\documentclass[11pt]{article}

\usepackage{amssymb, latexsym, amsmath}
\usepackage{amsfonts, graphics, amstext}
\usepackage{graphicx}
\usepackage{subcaption}
\usepackage{color}

\def\be{\begin{equation}}
\def\ee{\end{equation}}
\def\bea{\begin{eqnarray}}
\def\eea{\end{eqnarray}}
\def\beas{\begin{eqnarray*}}
\def\eeas{\end{eqnarray*}}

\newcommand{\prfe}{\hspace*{\fill} $\Box$

\smallskip \noindent}

\begin{document}

\sloppy

\newtheorem{theorem}{Theorem}[section]
\newtheorem{definition}[theorem]{Definition}
\newtheorem{proposition}[theorem]{Proposition}
\newtheorem{example}[theorem]{Example}
\newtheorem{remark}[theorem]{Remark}
\newtheorem{cor}[theorem]{Corollary}
\newtheorem{lemma}[theorem]{Lemma}

\renewcommand{\theequation}{\arabic{section}.\arabic{equation}}

\title{Existence of steady states of the massless Einstein-Vlasov system surrounding a Schwarzschild black hole}

\author{H{\aa}kan Andr\'{e}asson\\
        Mathematical Sciences\\
        Chalmers University of Technology\\
        University of Gothenburg\\
        SE-41296 Gothenburg, Sweden\\
        email: hand@chalmers.se}

\maketitle

\begin{abstract}
We show that there exist steady states of the massless Einstein-Vlasov system which surround a Schwarzschild black hole. The steady states are (thick) shells with finite mass and compact support. Furthermore we prove that an arbitrary number of shells, necessarily well separated, can surround the black hole. To our knowledge this is the first result of static self-gravitating solutions to any massless Einstein-matter system which surround a black hole. We also include a numerical investigation about the properties of the shells. 

\end{abstract}

\section{Introduction}
The Einstein-Vlasov system typically models self-gravitating particle ensembles such as galaxies or clusters of galaxies. The particles in the former case are stars and in the latter case they are galaxies. Clearly, the particles carry mass in these two situations. In this work we are instead interested in the case of massless particles, e.g. photons, and we show that there exist self-gravitating ensembles of massless particles with finite mass and compact support surrounding a Schwarzschild black hole. 
To put our result in context let us briefly review some related results. Existence of steady states to the Einstein-Vlasov system in the case of massive particles was first established in \cite{RR}. The steady states constructed in this work are spherically symmetric with a regular centre. Several simplifications and generalizations have since then been obtained and we refer to \cite{RaR} for a simplified and general approach, to \cite{A1} for the existence of highly relativistic static solutions and to \cite{AKR} for the existence of stationary solutions in the axisymmetric case. There are several other existence results and also results about the properties of the static solutions in the literature, and we refer to \cite{A3} for a review and to \cite{AFT1, AAL1, AAL2} for more recent results.

By relaxing the condition of a regular centre the case with a Schwarzschild black hole was considered in \cite{R1}, where the existence of massive static shells of Vlasov matter surrounding a black hole was shown. A different method leading to a similar result was more recently given by Jabiri in \cite{J}. For a fluid, the first result of a massive static shell surrounding a black hole was obtained in \cite{FHK}. If the matter model originates from quantum mechanics similar results need not be true as is for instance shown in \cite{FSY}, where the absence of static black hole solutions is shown for the Einstein-Dirac-Yang/Mills equations. It is argued in \cite{FSY} that a reason for the difference between the classical and the quantum mechanical case is that classical particles are prevented from falling into the black hole by the centrifugal barrier, whereas quantum particles can tunnel through this barrier. 

Solutions of the Einstein-Vlasov system can also model ensembles of massless particles, e.g. photons. 
The first mathematical study of the massless Einstein-Vlasov system is to our knowledge the work \cite{Ren} by Rendall, where the dynamics of cosmological solutions is investigated. Only more recently results about static solutions have been obtained. Akbarian and Choptuik constructed massless solutions with compact support numerically in \cite{AC}. An existence proof was obtained in \cite{AFT2}, where also a discussion about the relation to Wheeler's concept of geons is given. Gundlach studied the problem by numeric and analytic tools in \cite{G}. An important difference between the massive and massless case is that the existence of massless static solutions requires that the solutions are highly relativistic in the sense that the compactness ratio $2M/R$ is large. Here $M$ is the ADM mass of the solution and $R$ its (areal) radial support. It is known that $2M/R$ is always bounded by $8/9$, cf. \cite{A2}. (The classical result by Buchdahl \cite{Bu} does not apply in this case although the bound is the same.) Numerically it has been found that a necessary lower bound is roughly $2M/R>4/5$, cf. \cite{AC, AFT2, G}, for the existence of massless static solutions whereas no such lower bound is needed in the massive case. 

In the present work we combine the methods from \cite{R1} and \cite{AFT2}. We consider the case with a Schwarzschild black hole in the centre and we show that there exist static massless shells of Vlasov matter with compact support and finite mass which surround the black hole. Necessarily there is a gap between the black hole and the shell; the inner radius of the shell has to be larger than the radius of the photon sphere of the black hole. In our proof the shell is placed far away from the photon sphere. This is a technical condition. Numerically, we find that there are situations when the shell can be arbitrary close to the photon sphere, cf. Section 5. The shell solutions are highly relativistic in the sense that $2M/R$ is large. However, when the shell can be placed close to the photon sphere the ratio $2M/R$ is larger than, but close to, $2/3$. Hence, the presence of a black hole reduces the required lower bound of $2M/R$. Clearly, since the ratio $2M/R$ of the shell is larger than $2/3$ there is a photon sphere surrounding the shell in addition to the photon sphere which surrounds the black hole and which is situated between the black hole and the shell. 
Our result can be generalized to the case of an arbitrary number of shells. The resulting spacetime thus contains an arbitrary number of photon spheres. This seems to contradict the result in \cite{CG} which shows that only one photon sphere can appear in a static spacetime. However, the result in \cite{CG} does not apply in the case when the photon spheres are nested as in our case. 

We remark that our result holds also in the case when the black hole mass vanishes. However, the family of solutions obtained in this work is different from the family of solutions obtained in \cite{AFT2}. In the present situation we require the inner radius of the shell, $R_0$, to be sufficiently large, whereas $R_0$ is required to be sufficiently small in \cite{AFT2}. In fact, the compactness ratio $2M/R\to 8/9$ in \textit{both} the limits $R_0\to 0$ and $R_0\to\infty$. 

Let us finally mention that the linear massless Einstein-Vlasov system has been studied on a fixed black hole spacetime in \cite{ABJ}. The authors show that solutions to the linear Einstein-Vlasov system on a Kerr background satisfy a Morawetz estimate. Our result shows that an analogous result cannot hold for the nonlinear Einstein-Vlasov system. On the other hand, the main purpose of \cite{ABJ} is to understand perturbations of black hole spacetimes. The steady states we construct require compact configurations and the matter components cannot be made arbitrary small. Thus they should not be relevant when studying perturbations. 

The outline of the paper is as follows. In Section 2 we introduce the massless static Einstein-Vlasov system. In Section 3 we formulate the main results and in Section 4 we prove our main theorem. Section 5 is devoted to a numerical investigation of the properties of the solutions.

\section{The static Einstein\,-Vlasov system}

The metric of a static spherically symmetric spacetime takes 
the following form in Schwarzschild coordinates 
\begin{displaymath}
ds^{2}=-e^{2\mu(r)}dt^{2}+e^{2\lambda(r)}dr^{2}
+r^{2}(d\theta^{2}+\sin^{2}{\theta}d\varphi^{2}),
\end{displaymath}
where $r\geq 0,\,\theta\in [0,\pi],\,\varphi\in [0,2\pi].$
Asymptotic flatness is expressed by the boundary conditions
\begin{displaymath}
\lim_{r\rightarrow\infty}\lambda(r)=\lim_{r\rightarrow\infty}\mu(r)
=0. 
\end{displaymath}
We now formulate the static massless Einstein-Vlasov system. For an introduction to the Einstein-Vlasov system we refer to \cite{A3}, \cite{R3} and \cite{Ren}. 
Below we use units such that $c=G=1$ where $G$ is the gravitational constant and $c$ is the speed of light. 
The static massless Einstein-Vlasov system is given by the Einstein equations 
\begin{eqnarray}
&\displaystyle e^{-2\lambda}(2r\lambda_{r}-1)+1=8\pi r^2\rho,&\label{ee12}\\
&\displaystyle e^{-2\lambda}(2r\mu_{r}+1)-1=8\pi r^2
p,&\label{ee22}\\
&\displaystyle
\mu_{rr}+(\mu_{r}-\lambda_{r})(\mu_{r}+\frac{1}{r})= 8\pi
p_T e^{2\lambda},&\label{ee4}
\end{eqnarray}
together with the static Vlasov equation 
\begin{equation}
\frac{w}{\varepsilon}\partial_{r}f -(\mu_{r} \varepsilon-
\frac{L}{r^3\varepsilon})\partial_{w}f=0,\label{vlasov}
\end{equation}
where
\begin{equation*}
\varepsilon=\varepsilon(r,w,L)=\sqrt{w^{2}+L/r^{2}}.
\end{equation*}
The matter quantities are defined by
\begin{eqnarray}
\rho(r)&=&\frac{\pi}{r^{2}}
\int_{-\infty}^{\infty}\int_{0}^{\infty}\varepsilon(r,w,L) f(r,w,L)\;dLdw,\label{rho21}\\
p(r)&=&\frac{\pi}{r^{2}}\int_{-\infty}^{\infty}\int_{0}^{\infty}
\frac{w^{2}}{\varepsilon(r,w,L)}f(r,w,L)\;dLdw,\label{p21}\\
p_T(r)&=&\frac{\pi}{2r^{4}}\int_{-\infty}^{\infty}\int_{0}^{\infty}\frac{L}{\varepsilon(r,w,L)}f(r,w,L)\;
dLdw. 
\end{eqnarray}
The variables $w$ and $L$ can be thought of as the momentum in the radial direction 
and the square of the angular momentum respectively.

The matter quantities $\rho, p$ and $p_T$ are the energy density, the radial pressure and the tangential pressure respectively. 
The system of equations above are not independent and we study the reduced system (\ref{ee12})-(\ref{ee22}) together with (\ref{vlasov}) and (\ref{rho21})-(\ref{p21}). 
It is straightforward to show that a solution to the reduced system is a solution to the full system. 

Define $$E=e^{\mu}\varepsilon,$$ then the ansatz 
\begin{equation}
f(r,w,L)=\Phi(E,L), \label{ansatz}
\end{equation}
satisfies (\ref{vlasov}). 
By inserting this ansatz into (\ref{rho21})-(\ref{p21}) the system of equations reduce to a system where the metric coefficients $\mu$ and $\lambda$ alone are the unknowns. 
This has turned out to be an efficient method to construct static solutions and we will use this approach here. 
The following form of $\Phi$ will be used 
\begin{equation}
\Phi(E,L)=(E_0-E)^k_{+}(L-L_0)_{+}^l,\label{pol}
\end{equation}
where $l\geq 1/2,\,k\geq 0,\,L_0>0,\; E_0>0,$ and $x_{+}:=\max\{x,0\}$.
In the Newtonian case with $l=L_0=0,$ this ansatz leads
to steady states with a polytropic equation of state.  

The aim in this work is to show that static shells of Vlasov matter exist which surround a Schwarzschild black hole; in fact there can be arbitrary many shells separated by vacuum surrounding the black hole. 
To prove our result we construct highly compact shells, i.e., shells for which the compactness ratio 
\begin{equation}
\Gamma:=\sup \frac{2m(r)}{r},
\end{equation}
is large; roughly $\Gamma\geq \frac45$. Here $m$ is the Hawking mass defined for $r\geq 2M_0$ by 
\[
m(r)=M_0+\int_{2M_0}^r s^2\rho(s)\, ds, \;\; r\geq 2M_0.
\]
From \cite{A2} it always holds that $\Gamma<\frac89$. The result in \cite{A2} concerns steady states with a regular centre but it is straightforward to show that it holds also in the case with a Schwarzschild black hole at the centre. 

\begin{remark}Numerically we are able to construct solutions where the shell is close to the photon sphere of the black hole. For such solutions it turns out that $\Gamma$ is larger than, but close to, $2/3$, cf. Section 5. Hence, the presence of a black hole reduces the required lower bound of $\Gamma$. Indeed, recall that the numerical studies in the regular case indicates that the required lower bound is larger in that case, cf. \cite{AFT2}, \cite{AC} and \cite{G}. 
\end{remark}

If the inner radius of the shell is denoted by $R_0$, we show that for highly relativistic shells there is a radius $R_1$ such that $f(r,w,L)=0$ in an interval $[R_1,R_1+\epsilon], \epsilon>0$.  This fact makes it possible to glue a Schwarzschild solution at $r=R_1$ to the shell solution with support in $[R_0,R_1]$. If a Schwarzschild solution is not attached at $r=R_1$, then the ansatz (\ref{pol}) implies that Vlasov matter will occur again and there exists a radius $R_2$ such that $f>0$ for all $r>R_2$ and the solution is not asymptotically flat. This is a general feature of massless static solutions of the Einstein-Vlasov system obtained from an ansatz, cf. equation (\ref{ansatz}). In the massive case the situation is different and solutions generated by the ansatz (\ref{pol}) alone gives rise to compactly supported solutions. 

In the massive case the existence of shells surrounding a Schwarzschild black hole was settled in \cite{R3}. These shells are not highly relativistic. To construct highly relativistic shells for which $\Gamma$ is sufficiently large we adapt the method developed in \cite{A1}, which in turn was used to show existence of massless steady states with a regular centre in \cite{AFT2}. 

\section{Set up and main result}
Let $M_0>0$ be the mass of the black hole. In a vacuum region in the exterior of the black hole it holds that 
\[
e^{2\mu(r)}=1-\frac{2M_0}{r}. 
\]
Note that the ansatz (\ref{pol}) implies that $f=0$ whenever $E>E_0$. Accordingly we let $f=0$ in the interval $[2M_0,R_0],$ where $R_0$ is the largest root to the equation 
\begin{equation}\label{groundequation}
\big(1-\frac{2M_0}{r})\frac{L_0}{r^2}= E_0^2=1.
\end{equation}
Of course, we need a condition on $L_0$ which guarantees that the equation has real roots.

\begin{remark}The parameter $L_0$ can be removed. From above we see that by replacing $E_0$ by $\tilde{E_0}=E_0/\sqrt{L_0}$ we could consider the case $L_0=1$ and use $E_0$ as free parameter, cf. \cite{RV,G}. However, below we keep $L_0$ as free parameter and we fix $E_0=1$. 
\end{remark}

An elementary computation shows that the maximum value of the left hand side of (\ref{groundequation}) is 
\[
\frac{L_0}{27M_0^2},
\]
attained at $r=3M_0$. This radius corresponds to the radius of the photon sphere of the black hole. We fix $E_0=1$ and impose the condition that 
\[
L_0>27M_0^2=:L_*. 
\]
Equation (\ref{groundequation}) then has three real roots and we denote the largest root by $R_0$. 
To carry out the proof of our main result we will take $L_0$ large. We have the following result. 
\begin{lemma}\label{Walpha}
There exists a constant $C>0$, depending on $M_0$, such that 
\[
1-\frac{C}{\sqrt{L_0}}\leq \frac{R_0}{\sqrt{L_0}}\leq 1 \mbox{ as } L_0\to \infty. 
\]
\end{lemma}
\textbf{Proof: }The proof is a straightforward application of the solution formula for cubic equations. 
Indeed, equation (\ref{groundequation}) is equivalent to 
\[
r^3-L_0r+2M_0L_0=0,
\]
which we write as $r^3+pr+q=0$ where $p=-L_0$ and $q=2M_0L_0$. Set 
\[
D=(\frac{p}{3})^3+(\frac{q}{2})^2=L_0^2(-\frac{L_0}{27}+M_0^2). 
\]
Since we choose $L_0>27M_0^2$ it follows that $D<0$ and a standard result for cubic equations implies that the equation has three real roots given by
\[
r_1=u+v,\;\; r_{2,3}=\frac{u+v}{2}\pm \frac{u-v}{2}i\sqrt{3},
\]
where $u=(-q/2+i\sqrt{|D|})^{1/3}$ and $v=(-q/2-i\sqrt{|D|})^{1/3}$. Note here that $u+v$ and $(u-v)i$ are real. The largest of these roots is 
\[
R_0:=r_1=u+v=2(\frac{q^2}{4}+|D|)^{1/6}\cos{\frac{\alpha}{3}},
\]
where 
\[
\alpha=
\frac{\pi}{2}+\arctan{\frac{(q/2)}{\sqrt{|D|}}}.
\]
We have
\[
0\leq \frac{(q/2)}{\sqrt{|D|}}=\leq \frac{M_0}{\sqrt{L_0}\sqrt{\frac{1}{27}-\frac{M_0^2}{L_0}}}\leq \frac{C}{\sqrt{L_0}},
\]
where we used that $L_0-L_*>0$ since we consider large $L_0$. Hence there exists a constant $C>0$ such that 
\[
\frac{\pi}{2}\leq \alpha\leq \frac{\pi}{2}+\frac{C}{\sqrt{L_0}},
\]
for large $L_0$. This implies that 
\[
\frac{\sqrt{3}}{2}-\frac{C}{\sqrt{L_0}}\leq \cos{\frac{\alpha}{3}}\leq \frac{\sqrt{3}}{2},
\]
for some positive constant $C$. 
Moreover, we have 
\[
2(\frac{q^2}{4}+|D|)^{1/6}=2(M_0^2L_0^2+L_0^2(\frac{L_0}{27}-M_0^2))^{1/6}=\frac{2\sqrt{L_0}}{\sqrt{3}}. 
\]
Hence we conclude that for large $L_0$
\[
1-\frac{C}{\sqrt{L_0}}\leq \frac{R_0}{\sqrt{L_0}}\leq 1.
\]
\prfe

We now consider the Einstein-Vlasov system on the domain $r\geq R_0$ and we prescribe data for $\mu$ and $\lambda$ at $r=R_0$ by letting 
\[
e^{2\mu(R_0)}=e^{-2\lambda(R_0)}=1-\frac{2M_0}{R_0}. 
\]
Using results from previous works we can assume that there exists a solution to the Einstein-Vlasov system which exists on $[R_0,\infty[$ with the property that $\Gamma<8/9$ everywhere, cf. \cite{RR} for existence and \cite{A2} for the bound on $\Gamma$. We will show that if $R_0$ is sufficiently large, i.e. we take $L_0$ large, there is a radius $R_1>R_0$ such that the energy density and the pressure components vanish at $r=R_1$. The shell 
$[R_0,R_1]$ is thin in the sense that 
\begin{equation}\label{thin}
\frac{R_1}{R_0}\to 1 \mbox{as } R_0\to \infty. 
\end{equation} 
However, the difference $R_1-R_0$ does not need to decrease. Depending on the parameters in equation (\ref{pol}) the difference may even become unbounded, cf. Remark (\ref{remark-q}). Hence the shell is thin in the sense (\ref{thin}), which is different from the usual notion of thin shells in general relativity. 

When the distribution function $f$ has the form 
\begin{equation}
f(r,w,L)=\Phi(E,L),
\end{equation}
where $\Phi=0$ whenever $E>E_0$, 
the matter quantities $\rho$ and $p$ become
functionals of $\mu$, and we have 
\begin{eqnarray}
\rho &=& \frac{2\pi}{r^2}\int_{\frac{\sqrt{L_0}}{r}}^{E_0e^{-\mu}}\int_{L_0}^{r^2(s^2-1)}\Phi(e^{\mu}s,L)\frac{s^2}{\sqrt{s^2-1-\frac{L}{r^2}}}dLds,\label{rho}\\
p &=& \frac{2\pi}{r^2}\int_{\frac{\sqrt{L_0}}{r}}^{E_0e^{-\mu}}\int_{L_0}^{r^2(s^2-1)}\Phi(e^{\mu}s,L)\sqrt{s^2-1-\frac{L}{r^2}}dLds.\label{p}\\
\end{eqnarray}
Here we have kept the parameter $E_0$ but in what follows we use that $E_0=1$. 
By taking (\ref{pol}) for $\Phi$ these integrals can be computed explicitly 
in the cases when $k=0,1,2,...$ and $l=1/2,3/2,...$ 
Let $$\gamma=-\mu-\frac{1}{2}\log{\frac{L_0}{r^2}},$$ we then have the following lemma from \cite{AFT2}. 
\begin{lemma}Let $k=0,1,2,...$ and let $l=1/2,3/2,5/2,...$ then there are positive constants $\pi^{j}_{k,l},\, j=1,2,3$ such that when $\gamma\geq 0$
\begin{eqnarray}
\rho &=& \pi^{1}_{k,l}r^{2l}(\frac{L_0}{r^2})^{l+2}(e^{\gamma}-1)^{l+k+3/2}P_{l+5/2-k}(e^{\gamma}),\label{rhocon}\\
p &=& \pi^{2}_{k,l}r^{2l}(\frac{L_0}{r^2})^{l+2}(e^{\gamma}-1)^{l+k+5/2}P_{l+3/2-k}(e^{\gamma}),\label{pcon}
\end{eqnarray}
If $\gamma<0$ then all matter components vanish. 
Here $P_{n}(e^{\gamma})$ is a polynomial of degree $n$ and $P_n>0.$ 
\end{lemma}
In order to simplify the technical details we consider only the case $k=0$ and $l=1/2$ but we emphasize that our result holds more generally as in \cite{AFT2} and \cite{A1}. 

\begin{remark}\label{remark-q}
In the case $k=0$ and $l=1/2$ the support of the shell $[R_0,R_1]$ satisfy $R_1-R_0\leq C$, independently of $R_0$ as shown below, whereas for other values of the parameters we claim that 
\[
R_1-R_0\sim R_0^{\frac{q-2-2l}{q}},
\] 
where $q=k+l+5/2$, cf. Section 5. This relation is obtained by performing the analysis below in the general case, cf. \cite{A1}. 
\end{remark}
Let $t=e^{\gamma}$, we then have (with $k=0$ and $l=1/2$) 
\begin{equation}\label{rho1}
\rho(r)=\pi^2 r\big(\frac{L_0}{r^2}\big)^{5/2}(t-1)^2\big(\frac{3t^3+6t^2+4t+2}{15}\big),
\end{equation}
and
\begin{equation}\label{p1}
p(r)=\pi^2 r\big(\frac{L_0}{r^2}\big)^{5/2}(t-1)^3\big(\frac{3t^2+9t+8}{60}\big).
\end{equation}
Let $R_0$ be large and define 
\begin{equation}\label{delta}
\delta:=\big(\frac{1}{20\pi^3}\big)^{1/3}. 
\end{equation}
The argument below will be carried out in the interval $I:=[R_0,R_1]$ where $R_1<R_0+50\delta=:R_1^*$. Since $R_0$ will be taken large, $R_1/R_0$ is as close to $1$ as we wish. 
We now formulate the main results in this work. 
\begin{theorem}\label{thm-1}
Let $M_0\geq 0$ be the ADM mass of a Schwarzschild black hole. Then there exist static solutions with finite ADM mass to the massless spherically symmetric Einstein-Vlasov system surrounding the black hole. The matter components are supported on a finite interval $[R_0,R_1]$, where $R_0>3M_0$, and spacetime is asymptotically flat. 
\end{theorem}

\begin{remark}The arguments below lead to a solution for which $\mu(r)$ has a finite limit $\mu(\infty)$ as $r\to\infty$. In order to obtain an asymptotically flat solution we rescale by letting $\tilde{E_0}:=e^{\mu(\infty)}$ and $\tilde{\mu}(r):=\mu(r)-\mu(\infty)$. 
\end{remark}

\begin{remark}The regularity of the solution depends on the parameters $k$ and $l$, cf. equations (\ref{rhocon}) and (\ref{pcon}). For the values of $k$ and $l$ that we consider in this work the matter quantities are continuously differentiable. 
\end{remark}

\begin{remark}As mentioned in the introduction, note that our result holds also in the case when $M_0=0$ but that the family of solutions obtained in this work is different from the family of solutions obtained in \cite{AFT2}. In the present situation we require the inner radius $R_0$ to be large whereas in \cite{AFT2} $R_0$ is required to be small.  Clearly, when $M_0>0$ it is not possible to take $R_0$ small since necessarily $R_0>2M_0$. Both families share the property that $\Gamma\to 8/9$ in the extreme limits, i.e., when $R_0\to 0$ as in \cite{AFT2} or when $R_0\to \infty$ as in the present case. 
\end{remark}

\begin{remark}Having a black hole with one shell surrounding it, we can start from this solution and add another shell with the strategy in the proof. Hence, our result implies that an arbitrary number of shells can surround the black hole. 
\end{remark}

In view of the discussion in Section 2, Theorem \ref{thm-1} is a consequence of the following result. 
\begin{theorem}\label{thm-2}
Consider a static solution to the massless Einstein-Vlasov system, corresponding to the ansatz (\ref{pol}) with $k=0$ and $l=1/2$, with data given at $r=R_0$. For $R_0$ sufficiently large, there exists $R_1\leq R_0+50\delta$ such that the matter components are supported in the interval $[R_0,R_1]$ and vanish at $r=R_1$. Here $\delta$ is given by (\ref{delta}). Furthermore, $\Gamma\to\frac89$ as $R_0\to\infty$. 
\end{theorem}

\section{Proof of main result}
The proof of our main result will follow from a chain of lemmas. 

\textit{Proof of Theorem \ref{thm-2}}. 
First we establish convenient formulas for the matter terms. Although $L_0$ is our free parameter, we will instead use $R_0$ as free parameter since $R_0\to\infty$ as $L_0 \to\infty$ in view of Lemma \ref{Walpha}. Next we note that $R_1^*/R_0\leq 1+C/R_0$. Hence, by Lemma \ref{Walpha} we have for any $r\in [R_0,R_1^*]$ 
\begin{equation}\label{approx1}
1-\alpha(R_0)\leq{\frac{L_0}{r^2}}\leq 1+\alpha(R_0),
\end{equation}
where $\alpha(R_0)\geq 0$ and $\alpha(R_0)\to 0$ as $R_0\to \infty$. 
By the definition of $R_0$ we have that $\gamma(R_0)=0$. Moreover,
\begin{equation}\label{gammaprime}
\gamma'(r)=-\mu'(r)+\frac{1}{r}. 
\end{equation}
Now, since $\rho=p=0$ and $m(r)=M_0$ for $r\leq R_0$, we have that that $\mu'(R_0)<1/R_0$. Here we used that $M_0/R_0<1/3$ and that 
\[
e^{2\lambda(r)}=\frac{1}{1-\frac{2m(r)}{r}}.
\]
The last relation is a consequence of the Einstein equation (\ref{ee12}). 
This implies that $\gamma'(R_0)>0$. 
Thus $\gamma(r)>0$ in a right neighborhood of $R_0$ and the aim is to show that there exists $R_1<R_1^*$ such that $\gamma(R_1)=0$. Hence $\gamma>0$ on the interior of the closed interval $I$. An upper bound on $\gamma$ follows from (\ref{gammaprime}). We have since $\mu'(r)\geq 0$, 
\[
\gamma(r)=\gamma(r)-\gamma(R_0)\leq \log\frac{r}{R_0}=\log(1+\frac{r-R_0}{R_0}),
\]
which implies that $0\leq \gamma\leq \alpha(R_0)$ on $I$ if $R_0$ is sufficiently large. Hence for $r\in I$ 
\[
\gamma(r)\leq e^{\gamma(r)}-1\leq \gamma(r)(1+\alpha(R_0)). 
\]
In particular this implies that (recall $t=e^{\gamma}$) 
\[
1\leq \frac{3t^3+6t^2+4t+2}{15}\leq 1+\alpha(R_0),
\]
and similarly for the corresponding polynomial in $p(r)$. 
Putting these estimates together we conclude that 
\[
\pi^2 r\gamma^2(r)(1-\alpha(R_0))\leq \rho(r)\leq \pi^2 r\gamma^2(r)(1+\alpha(R_0)),
\]
and similarly for $p(r)$. Since for all arguments below it is sufficient to have a lower and an upper bound on $\rho$ and $p$ we assume for simplicity that 
\begin{equation}\label{rho2}
\rho(r)=\pi^2 r\gamma^2(r),
\end{equation}
and 
\begin{equation}\label{p2}
p(r)=\frac{\pi^2}{3} r \gamma^3(r),
\end{equation}
for $r\in I$. 

\begin{lemma}\label{lemma3}
Let $\delta$ be as above. Then 
\[
\gamma'(r)\geq \frac{1}{2r}\; \mbox{for } r\in [R_0,R_0+\delta]=:I_1. 
\]
\end{lemma}
\textit{Proof: }We have by the mean value theorem that for any $\sigma\leq \delta$ 
\[
\gamma(R_0+\sigma)=\gamma(R_0+\sigma)-\gamma(R_0)=\sigma \gamma'(\xi)\leq \frac{\delta}{R_0},
\]
where $\xi\in [R_0,R_0+\sigma]$, since $\gamma(r)\leq 1/r$. Hence, 
\[
\rho(r)\leq \pi^2 r\frac{\delta^2}{R_0^2}\; \mbox{ on } I_1. 
\]
We get for $\sigma\leq \delta$ 
\[
m(R_0+\sigma)\leq M_0+\frac{4\pi^3\delta^2}{R_0^2}\int_{R_0}^{R_0+\sigma}\eta^3 \, d\eta\leq M_0+\frac{4\pi^3\delta^2(R_0+\sigma)^3}{R_0^2}\sigma.
\]
By taking $R_0$ large we thus obtain
\[
\frac{m(R_0+\sigma)}{R_0+\sigma}\leq 4\pi^3\delta^2\sigma +\alpha(R_0)\leq 4\pi^3\delta^3+\alpha(R_0)\leq \frac15 +\alpha(R_0).
\]
This implies that for $r\in I_1$
\[
\frac{m(r)}{r}e^{2\lambda(r)}\leq \frac13+\alpha(R_0).
\]
Moreover, we have for $r\in I_1$ that 
\[
4\pi r^2 p(r)e^{2\lambda(r)}\leq 4\pi r^2 p(r)(\frac53+\alpha(R_0))\leq \frac{4\pi^3\delta^3}{3}(\frac53+\alpha(R_0))\leq \frac19+\alpha(R_0).
\]
Hence, for $r\in I_1$ we get for sufficiently large $R_0$
\[
\mu'(r)\leq (\frac13+\frac19+\alpha(R_0))\frac{1}{r}\leq \frac{1}{2r}.
\]
This completes the proof of the lemma. 
\prfe
The lemma implies that for $\sigma\leq\delta$ 
\[
\gamma(R_0+\sigma)\geq \gamma(R_0)+\sigma\inf_{0\leq s\leq\sigma} \gamma'(R_0+s)\geq 
\frac{\sigma}{2(R_0+\sigma)}.
\]
Let 
\[
\sigma_*:=\delta,
\]
and define
\[
\gamma_*=\frac{\sigma_*}{2(R_0+\sigma_*)}. 
\]
\begin{remark}
The notation $\sigma_*$ is in this case superfluous but for general parameter values it is motivated, cf. \cite{A1}. 
\end{remark}

Clearly we have that
\[
\gamma(R_0+\sigma_*)\geq \gamma_*.
\]
The result in the following lemma shows that $\gamma$ will reach the $\gamma_*$ level again at a larger radius. 
\begin{lemma}\label{lemma4}
There exists a radius $r_2>R_0+\sigma_*$ such that $\gamma(r_2)=\gamma_*$ and such that $\gamma(r)>\gamma_*$ for $R_0+\sigma_*<r<r_2$. Moreover, $r_2\leq R_0+11\delta$. 
\end{lemma}
\textit{Proof: }Since $\gamma(R_0+\sigma_{*})\geq\gamma_{*},$ and since Lemma \ref{lemma3}  gives that $\gamma'(R_0+\sigma_{*})>0,$ the radius $r_2$ must be strictly greater than $R_0+\sigma_{*}.$ Let $[R_0+\sigma_{*},R_0+\sigma_{*}+\Delta],$ for some $0<\Delta<10 \delta,$ be the smallest interval such that $\gamma\geq\gamma_{*}$ on the interval. We will show that there in fact $\gamma(R_0+\sigma_*+\Delta)=\gamma_*$. 
We have from (\ref{rho2}) 
\begin{eqnarray}\label{msigma}
m(R_0+\sigma_*+\Delta)&\geq& 4\pi^3\int_{R_0+\sigma_{*}}^{R_0+\sigma_{*}+\Delta}\frac{\sigma_{*}^2}{4(R_0+\sigma_{*})^2}r^3\, dr\nonumber \\
&\geq&\pi^3\sigma_*^2(R_0+\sigma_*)\Delta.
\label{msigma}
\end{eqnarray}
Hence, 
\[
\frac{m(R_0+\sigma_{*}+\Delta)}{R_0+\sigma_{*}+\Delta}\geq \frac{\pi^3\sigma_{*}^2(R_0+\sigma_*)}{R_0+\sigma_{*}+\Delta}\Delta\geq \frac{9}{10} \pi^3\sigma_{*}^2\Delta,
\]
where we used that $R_0\geq 100\delta$, so that $(R_0+\sigma_*)/(R_0+\sigma_*+\Delta)\geq 9/10$, since we consider large $R_0$. 
Since for $r\geq R_0$ necessarily 
\[
\frac{m(r)}{r}<\frac49,
\]
we conclude that there is a $\Delta$ such that 
\[
\Delta\leq \frac{40}{81\pi^3 \sigma_*^2}=\frac{40}{81\pi^3 \delta^2}=\frac{40\cdot 20^{2/3}}{81\pi}\leq 10\delta, 
\]
with the property that $\gamma(R_0+\sigma_*+\Delta)=\gamma_*$, since otherwise we obtain a contradiction. 
\prfe
 Next we show an important property of the solution at the radius $r=r_2$ when $R_0$ is sufficiently large. \\
\begin{lemma}\label{lemma5}
Let $r_2$ be as above. If $R_0$ is sufficiently large then 
\[
\frac{m(r_2)}{r_2}\geq \frac25. 
\]
\end{lemma}
\textit{Proof: }
We consider the fundamental equation ($10$) in \cite{A0}. In our case with a black hole at the center it takes the form 
\begin{equation}
(\frac{m(r)}{r^2}+4\pi rp(r))e^{\mu(r)+\lambda(r)}-\frac{M_0}{R_0^2}=\frac{1}{r^2}\int_{R_0}^r 4\pi\eta^2 e^{\mu+\lambda}(\rho+p+2p_T)d\eta.\label{fundamentaleq}
\end{equation}
Here we used that $\mu(R_0)+\lambda(R_0)=0$ and that $p(R_0)=0$. This equation is a consequence of the generalized Oppenheimer-Tolman-Volkov equation. In the present massless case we have $p+2p_T=\rho$, so that $\rho+p+2p_T=2\rho$. If we take $r=r_2$ we then get 
\begin{eqnarray}
\frac{m(r_2)}{r_2}e^{(\mu+\lambda)(r_2)}&\geq&\frac{8\pi}{r_2}\int_{R_0}^{r_2} \eta^2 e^{\mu+\lambda}\rho\,d\eta-4\pi r_2^2pe^{(\mu+\lambda)(r_2)}.\label{Mr2inequality} 
\end{eqnarray}
Here we dropped the term involving $M_0$ due to sign. Note that it becomes arbitrary small for $R_0$ large. 
Next we write 
\begin{eqnarray}
\int_{R_0}^{r_2}4\pi\eta\rho e^{\lambda}\,d\eta&=&\int_{R_0}^{r_2}(-\frac{d}{dr}e^{-\lambda})d\eta+\int_{R_0}^{r_2}\frac{me^{\lambda}}{\eta^2}d\eta\nonumber\\
&\geq& \sqrt{1-\frac{2M_0}{R_0}}-\sqrt{1-\frac{2m(r_2)}{r_2}}\nonumber\\
&=& 1-\sqrt{1-\frac{2m(r_2)}{r_2}}-\big(1-\sqrt{1-\frac{2M_0}{R_0}}\big)\nonumber \\
&=&\frac{2m(r_2)}{r_2(1+\sqrt{1-\frac{2m(r_2)}{r_2}})}-\alpha(R_0).\label{fundcomp}
\end{eqnarray}
In the inequality above we dropped the second term due to sign. We note that this term is as small as we wish for a relatively thin shell. The reason we point this out is that the chain of \textit{inequalities} in this paragraph is close to a chain of \textit{equalities} for very thin shells. This is essential to understand why for a thin shell $\Gamma$ approaches the limit $8/9$.

From (\ref{Mr2inequality}) we obtain using that $\mu$ is increasing
\begin{eqnarray}Y
\frac{m(r_2)}{r_2}e^{(\mu+\lambda)(r_2)}&\geq& \frac{8\pi}{r_2}\int_{R_0}^{r_2}\eta^2 e^{\mu+\lambda}\rho\,d\eta -4\pi r_2^2p(r_2)e^{(\mu+\lambda)(r_2)}\nonumber\\
&\geq& 8\pi e^{\mu(R_0)}\frac{R_0}{r_2}
\int_{R_0}^{r_2}\eta e^{\lambda}\rho\,d\eta-4\pi r_2^2 p(r_2)e^{(\mu+\lambda)(r_2)}.\nonumber\ \label{Mr2inequalityz} 
\end{eqnarray}
Let us introduce the notation $P:=4\pi r_2^2 p(r_2)$. Using (\ref{fundcomp}) we get
\begin{eqnarray}
\frac{m(r_2)}{r_2}e^{(\mu+\lambda)(r_2)}&\geq&\frac{e^{\mu(R_0)}R_0}{r_2}\frac{4m(r_2)}{r_2(1+\sqrt{1-\frac{2m(r_2)}{r_2}})}\nonumber\\
& &-Pe^{(\mu+\lambda)(r_2)}-\alpha(R_0).
\end{eqnarray}
Letting
\[
Y:=\frac{m(r_2)}{r_2},
\]
we obtain by using that $e^{-\lambda(r_2)}=\sqrt{1-2Y(r_2)}$ 
\begin{equation}
Y\geq e^{\mu(R_0)-\mu(r_2)}\left(\frac{R_0}{r_2}\right)\frac{4Y\sqrt{1-2Y}}{1+\sqrt{1-2Y}}-P-\alpha(R_0)e^{-(\mu+\lambda)(r_2)}. \label{eqnbuchdahl}
\end{equation}
Next we show a couple of properties of the solutions that we need to proceed with the argument. 
From Lemma 4 we have that $\gamma$ approaches $\gamma_{*}$ from above and therefore $\gamma'(r_2)\leq 0,$ which implies that 
\begin{equation}\label{muprimeatr2}
\mu'(r_2)\geq \frac{1}{r_2},
\end{equation}
in view of (\ref{gammaprime}). 
Furthermore we have from (\ref{p2}) that
\begin{equation}\label{p-star}
p(r_2)=\frac{\pi^2}{3} r_2 \gamma_*^3=\frac{\pi^2}{3} r_2\frac{\delta^3}{8(R_0+\delta)^3}.\nonumber
\end{equation}
Now recall that 
\[
\mu'=(\frac{m}{r^2}+4\pi rp)e^{2\lambda}.
\]
We will show that $Y=m(r_2)/r_2\geq 1/4$. 
Assume the contrary, i.e., assume 
\begin{equation}\label{contrary}
Y<\frac14,
\end{equation}
then 
\[
e^{2\lambda(r_2)}=\frac{1}{1-2Y}<2,
\]
and thus
\begin{equation}\label{p2r2}
4\pi r_2^2p(r_2)e^{2\lambda(r_2)}\leq \frac{\pi^3}{3} r_2^3\frac{\delta^3}{(R_0+\delta)^3}=
\frac{1}{60}\frac{r_2^3}{(R_0+\delta)^3}\leq\frac{1}{50}.
\end{equation}
Here we used that $\delta^3=1/20\pi^3$ and that $r_2\leq R_0+11\delta$ so that 
\[
\frac{r_2^3}{(R_0+\delta)^3}\leq\frac{6}{5} \mbox{ for }R_0 \mbox{ large}.
\]  
Hence if (\ref{contrary}) holds then 
\[
\mu'(r_2)=(\frac{m(r_2)}{r_2^2}+4\pi r_2p(r_2))e^{2\lambda(r_2)}\leq (\frac{1}{2}+\frac{1}{50})
\frac{1}{r_2},
\]
which is a contradiction to (\ref{muprimeatr2}) and we obtain that 
\[
Y\geq \frac14.
\]
Let us next consider the difference $\mu(R_0)-\mu(r_2).$ We have  
\[
\mu(r_2)-\mu(R_0)=\int_{R_0}^{r_2} (\frac{m}{\eta^2}+4\pi\eta p)e^{2\lambda}\,d\eta. 
\]
Since $\gamma'(r)\leq 1/r$ we have for $R_0\leq r\leq r_2\leq R_0+11\delta$
\[
\gamma(r)\leq \log{\frac{r}{R_0}}\leq \frac{11\delta}{R_0},
\]
and we get by using (\ref{p2}), and that $e^{2\lambda}\leq 9$ 
\[
4\pi r^2 p(r)e^{2\lambda(r)}\leq 12\pi^3 r^3 \gamma^3(r)\leq \frac{3}{5}\big(\frac{r}{R_0}\big)^3 11^3\leq C.
\] 
Hence, since $m/r\leq 4/9$ we get 
\[
\int_{R_0}^{r_2} (\frac{m}{\eta^2}+4\pi\eta p)e^{2\lambda}\,d\eta\leq \int_{R_0}^{r_2} \frac{C}{\eta}\,d\eta\leq C\log{\frac{r_2}{R_0}}\leq \frac{C}{R_0}.
\]
As a consequence we obtain that 
\[
1-\alpha(R_0)\leq e^{\mu(R_0)-\mu(r_2)}\left(\frac{R_0}{r_2}\right)\leq 1+\alpha(R_0).
\]
From this estimate we also have that $e^{-(\mu+\lambda)(r_2)}$ is bounded independently of $R_0$ and using that $Y\geq 1/4$ we can therefore write (\ref{eqnbuchdahl}) as
\begin{equation}\label{fundamentalGamma}
1\geq \frac{4\sqrt{1-2Y}}{1+\sqrt{1-2Y}}-4P-\alpha(R_0). 
\end{equation}
Let us introduce the notation $B=4P+\alpha(R_0)$. We can assume that $B\geq 0$ since otherwise this term can be dropped. Multiplying (\ref{fundamentalGamma}) by 
\[
1+\sqrt{1-2Y},
\]
and then squaring both sides we obtain 
\begin{equation}\label{fundamentalGamma2}
Y\geq \frac49-\frac{B\sqrt{1-2Y}(1+\sqrt{1-2Y})}{3},
\end{equation}
where we dropped the term involving $B^2$ due to sign. Using that $Y\geq 1/4$ we estimate
\[
\frac{\sqrt{1-2Y}(1+\sqrt{1-2Y})}{3}\leq \frac{1}{3\sqrt{2}}(1+\frac{1}{\sqrt{2}})\leq \frac{9}{20}. 
\]
Taking $R_0$ large we derive from (\ref{fundamentalGamma2}) the inequality
\[
Y\geq \frac49-2P.
\]
The estimate (\ref{p2r2}) can now be used noting that $e^{2\lambda(r_2)}\geq 2$ and we obtain 
\[
Y\geq \frac49-\frac{1}{50}\geq \frac25. 
\]

\prfe
\begin{remark}The proof above is very similar to the corresponding proof in \cite{A1}. However, for shells for which the inner radius $R_0\to 0$, as in \cite{A1}, the pressure term $P\to 0$ and one can conclude from the argument in the lemma above that as $R_0\to 0$ the compactness ratio $\Gamma\to 8/9$. In the present case the situation is slightly different. However, as soon as we know that there is a radius $R_1$ such that $\rho(R_1)=p(R_1)=0$, and such that $R_1/R_0\to 1$ as $R_0\to\infty$, then we can use the argument above, with $r_2$ replaced by $R_1$, to show that the compactness ratio $\Gamma\to \frac89$ as $R_0\to \infty$. This shows the last claim in Theorem \ref{thm-2}. 
\end{remark}
Inspired by an idea of T. Makino introduced in \cite{Ma}, we show that $\gamma$ necessarily must vanish close to the point $r_2$ if $R_0$ is sufficiently large. 
Let $$x:=\frac{m(r)}{r\gamma(r)}.$$
Using that $m'(r)=4\pi r^2\rho,$ and that $\mu'(r)=(m/r^2+4\pi r)e^{2\lambda}$ it follows that 
\[
rx'=\frac{4\pi r^2\rho}{\gamma}-x+\frac{x^2}{1-2\gamma x}-\frac{x}{\gamma}+
\frac{4\pi xr^2 p(r)}{\gamma (1-2\gamma x)}.
\]
In our case $r>R_0$ and $\gamma>0$ and we will show that $\gamma(r)=0$ for some $R_1\in I$.  Since $\gamma>0$ and $\rho,p\geq 0$ the first term and the last term can be dropped and we have 
\begin{equation}
rx'\geq -x+\frac{x^2}{1-2\gamma x}-\frac{x}{\gamma}=\frac{x^2}{3(1-2\gamma x)}-x+\frac{2x^2}{3(1-2\gamma x)}-\frac{x}{\gamma}. 
\end{equation}
Take $R_0$ sufficiently large so that $m(r_2)/r_2\geq 2/5$ by Lemma \ref{lemma5}. Let 
$r\in [r_2,16r_2/15],$ then since $m$ is increasing in $r$ we get 
\[
\frac{m(r)}{r}\geq\frac{m(r_2)}{r}=\frac{r_2}{r}\frac{m(r_2)}{r_2}\geq\frac{15}{16}\cdot\frac{2}{5}=\frac{3}{8}.\]
Now by the definition of $x$ it follows that 
$$\frac{x}{1-2\gamma x}=\frac{m}{\gamma r}e^{2\lambda}=\frac{m}{\gamma r(1-2m/r)}\geq\frac{3}{2\gamma}, \mbox{ when } \frac{m}{r}\geq\frac{3}{8}.$$
Thus on $[r_2,16r_2/15],$
$$\frac{2x^2}{3(1-2\gamma x)}-\frac{x}{\gamma}\geq 0,$$
so that on this interval 
\begin{equation}
rx'\geq\frac{x^2}{3(1-2\gamma x)}-x\geq\frac{4}{3}x^2-x,\label{makino} 
\end{equation}
where we used that 
$$\frac{1}{1-2\gamma x}=\frac{1}{1-2m/r}\geq 4 \mbox{ when } \frac{m}{r}\geq\frac{3}{8}.$$
The upper bound 
\[
\gamma\leq \log(1+\frac{r-R_0}{R_0})
\] 
implies that 
\begin{equation}\label{x2}
x(r_2)=\frac{m(r_2)}{r_2\gamma(r_2)}\geq \frac25\frac{R_0}{r_2-R_0}.
\end{equation}
In particular 
$$\frac{x(r_2)}{x(r_2)-3/4}\leq\frac{16}{15},$$
for $R_0$ large. 
Solving (\ref{makino}) yields 
$$x(r)\geq\frac34\left(1-\frac{r(4x(r_2)/3-1)}{4r_2x(r_2)/3}\right)^{-1}, \mbox{ on } r\in [r_2,16r_2/15),$$
and we get that $x(r)\to\infty$ as $r\to R_1,$ where
\begin{equation}
R_1\leq r_2\frac{x(r_2)}{x(r_2)-3/4}.\label{R1}
\end{equation}
In view of (\ref{x2}) we estimate
\[
R_1\leq r_2+\frac{\frac{15}{8}(r_2-R_0)r_2}{R_0-\frac{15}{8}(r_2-R_0)}.
\]
For sufficiently large $R_0$ we have $R_0/10\geq \frac{15}{8}(r_2-R_0)\geq 0$, recall that $r_2-R_0\leq 11\delta$, and we obtain
\[
R_1\leq  r_2+\frac{\frac{25}{12}(r_2-R_0)r_2}{R_0}\leq r_2+3(r_2-R_0)\leq R_0+50\delta. 
\]
This completes the proof of the theorem.
\prfe

\section{Numerical results}
In this section we present some numerical results. In the analytic proof we required $L_0$, or equivalently $R_0$, to be large. One aim is to investigate for which $L_0$ solutions can be obtained numerically. Recall that necessarily $L_0>L_*$, and one question is if solutions can be constructed for $L_0$ values arbitrary close to $L_*$, i.e., if the shell can occur arbitrary close to the photon sphere of the black hole. It turns out that this depends on the parameters and on the mass of the black hole, but for a large class of solutions it is possible. Hence, the condition in the proof that $L_0$ is large is mainly a technical condition. Moreover, based on our numerical findings together with the arguments in the proof, we conjecture that the compactness ratio for any shell surrounding a black hole satisfies $\Gamma>2/3$. Note that $2M_0/r=2/3$ at the radius of the photon sphere. This study is presented in Subsection 5.1. 

Another aim is to numerically confirm the claim in Remark (\ref{remark-q}) 
that the support $[R_1,R_0]$ of the solution satisfies 
\[
R_1-R_0\sim R_0^{\frac{q-2-2l}{q}},
\]
where $q=k+l+5/2$. Hence, the thickness of the shells depend on the parameter values $k$ and $l$. This investigation is presented in Subsection 5.2. 

In Subsection 5.3 we construct solutions with several shells. We do this by using two different strategies. From the set up in Section 2 we know that having a compactly supported solution with ADM mass $M_0$, where $M_0$ in this case is the total mass of the black hole and the shell(s) surrounding the black hole, we can choose $L_0>L_*$ (sufficiently large) and find the largest root $R_0$ to equation (\ref{groundequation}). We can then pose data at $r=R_0$ and numerically construct an additional shell. This strategy thus follows the analytic proof in the previous section. 

The second strategy is different in the sense that it does not follow the strategy of the proof. We start with a black hole and then we use \textit{one} ansatz function for the shells, in particular we do not change $L_0$ for the different shells. In this way a solution consisting of several shells, separated by vacuum, is generated by solving equation (\ref{groundequation}) only once. 
After a number of shells there is no longer a vacuum region separating the neighboring shells and a Schwarzschild solution has to be glued before this happens. 
The solution obtained in this way is very similar to the multi-peak solutions obtained in the massive case \cite{AR}. In the latter case there is however no need to glue a Schwarzschild solution since the massive solutions have compact support. 

\subsection{A black hole with one shell}
\begin{figure}
\begin{center}
\includegraphics[width=0.4\textwidth]{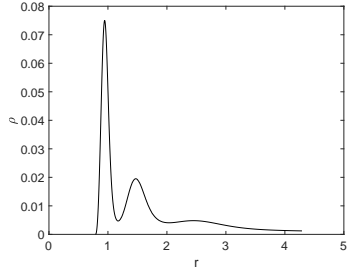}
\end{center}
\caption{Not a proper shell ($k=0, l=1/2$).}\label{naps}
\label{fig1}
\end{figure}

\begin{figure}
\begin{center}
\includegraphics[width=0.4\textwidth]{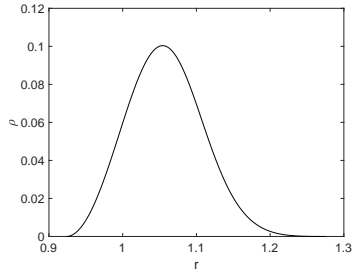}
\end{center}
\caption{A proper shell ($k=0, l=1/2$). $L_0=1.86, \Gamma=0.74$ and $r_*=0.75$.}\label{scttps}
\label{fig2}
\end{figure}

First we compute solutions for the parameter values $k=0$ and $l=1/2$ used in the proof. We have chosen the mass of the black hole to be  $M_0=0.25$ so that $L_*=27M_0^2=27/16\approx 1.69$. First we try to put the shell close to the photon sphere of the black hole by choosing $L_0=1.7$. However, as Figure \ref{fig1} shows there is no vacuum region after the first peak (and thus there is no proper shell) and it is not possible to obtain an asymptotically flat solution. (In all figures the energy density $\rho$ is displayed on the vertical axis and the area radius $r$ on the horizontal axis.) The choice $L_0=1.86$  gives on the other hand rise to a vacuum region after the first peak and an asymptotically flat solution is obtained. The photon sphere is located at $r=3M_0=0.75$, and the inner radius of the shell is located at $R_0=0.92$. The compactness ratio $\Gamma$ of the shell is $0.74$. The shell is depicted in Figure \ref{fig2}. The radius of the photon sphere preceding the shell is denoted by $r_*$. 

\begin{figure}
\begin{center}
\includegraphics[width=0.4\textwidth]{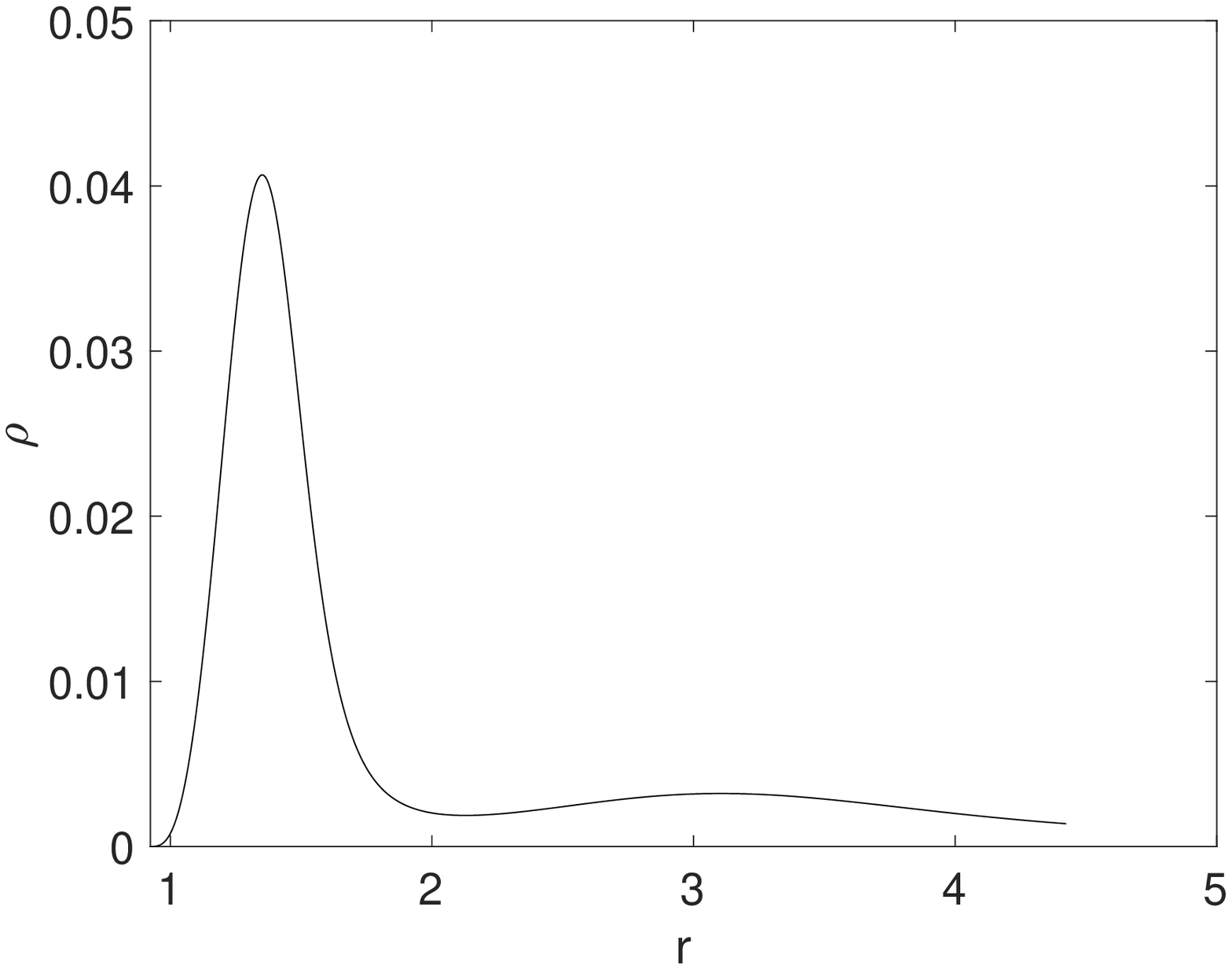}
\end{center}
\caption{Not a proper shell ($k=1, l=1/2$).}\label{naps}
\label{fig3}
\end{figure}

\begin{figure}
\begin{center}
\includegraphics[width=0.4\textwidth]{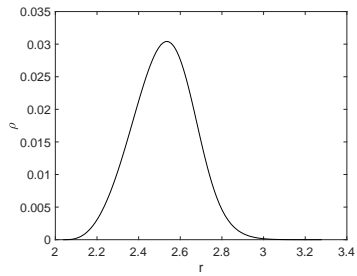}
\end{center}
\caption{A proper shell ($k=1, l=1/2$). $L_0=5.5, \Gamma=0.8$ and $r_*=0.75$.}
\label{fig4}
\end{figure}

If we increase the parameter value $k$, to $k=1$ instead of $k=0$, we have to increase $L_0$ to obtain a proper shell, i.e., the shell must be placed further away from the photon sphere of the black hole. Indeed, in Figure \ref{fig3} we have used the same parameter values as in Figure \ref{fig2} with the only change that $k=1$ instead of $k=0$. We see that in  this case a proper shell is not obtained. We have to increase $L_0$ to $L_0=5.5$ in order to get a proper shell. This case as shown in Figure \ref{fig4}. The inner radius is located at $R_0=2.04$ and the compactness ratio is $\Gamma=0.80 $. Hence, the parameter $k$ has a considerable influence on how close to the photon sphere the shell can be placed. We point out that the parameter $l$ has a similar impact but we have not made a systematic study of the dependence on $k$ and $l$. 

If we increase the mass of the black hole the situation changes and the shell can be placed closer to the photon sphere. Again this depends on the parameter values of $k$ and $l$, but generally the shells can be placed closer to the photon sphere when $M_0$ is larger. If $M_0$ is taken sufficiently large the shell can be placed arbitrary close to the photon sphere. This results in a low value of the compactness ratio $\Gamma$. Since $2M_0/r=2/3$ at the location of the photon sphere, we conjecture that $\Gamma>2/3$ for any shell surrounding the black hole, where low values are attained for shells which are located very close to the photon sphere. Recall from the proof that as the inner radius $R_0\to\infty$, the compactness ratio $\Gamma\to 8/9$ (a fact that is confirmed numerically in Figures \ref{fig:sfig3ch2}, \ref{fig:sfig3ch3} and \ref{fig:sfig3ch4}), so that morally it is advantageous to keep $R_0$ small in order to get a low value of $\Gamma$. 

\begin{figure}
\begin{center}
\includegraphics[width=0.4\textwidth]{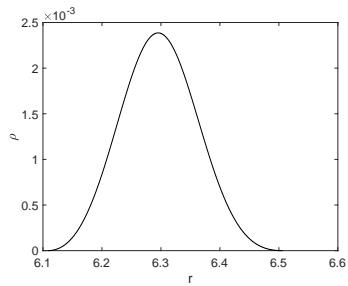}
\end{center}
\caption{A proper shell ($k=0, l=1/2$). $L_0=108.1, \Gamma=0.68$ and $r_*=6.0$.}\label{aps2}
\label{fig5}
\end{figure}

We choose $M_0=2.0$, which implies that $L_*=27M_0^2=108$. The photon sphere is located at $r=3M_0=6.0$. We find that the choice $L_0=108.1$ gives rise to a proper shell which is located very close to the photon sphere, at $R_0=6.1$. The shell is depicted in Figure  \ref{fig5}, where $k=0$ and $l=1/2$. The compactness ratio $\Gamma$ for the shell is merely $0.68$. If $k$ is increased to $k=1$, then we need to increase $L_0$, to $L_0=113$, in order to obtain a proper shell, which results in a shell with an inner radius of $R_0=6.9$, and a compactness ratio $\Gamma=0.73$.

\subsection{The thickness of the shells}
\begin{figure}
\begin{subfigure}{.4\textwidth}
  \centering
  \includegraphics[width=.8\linewidth]{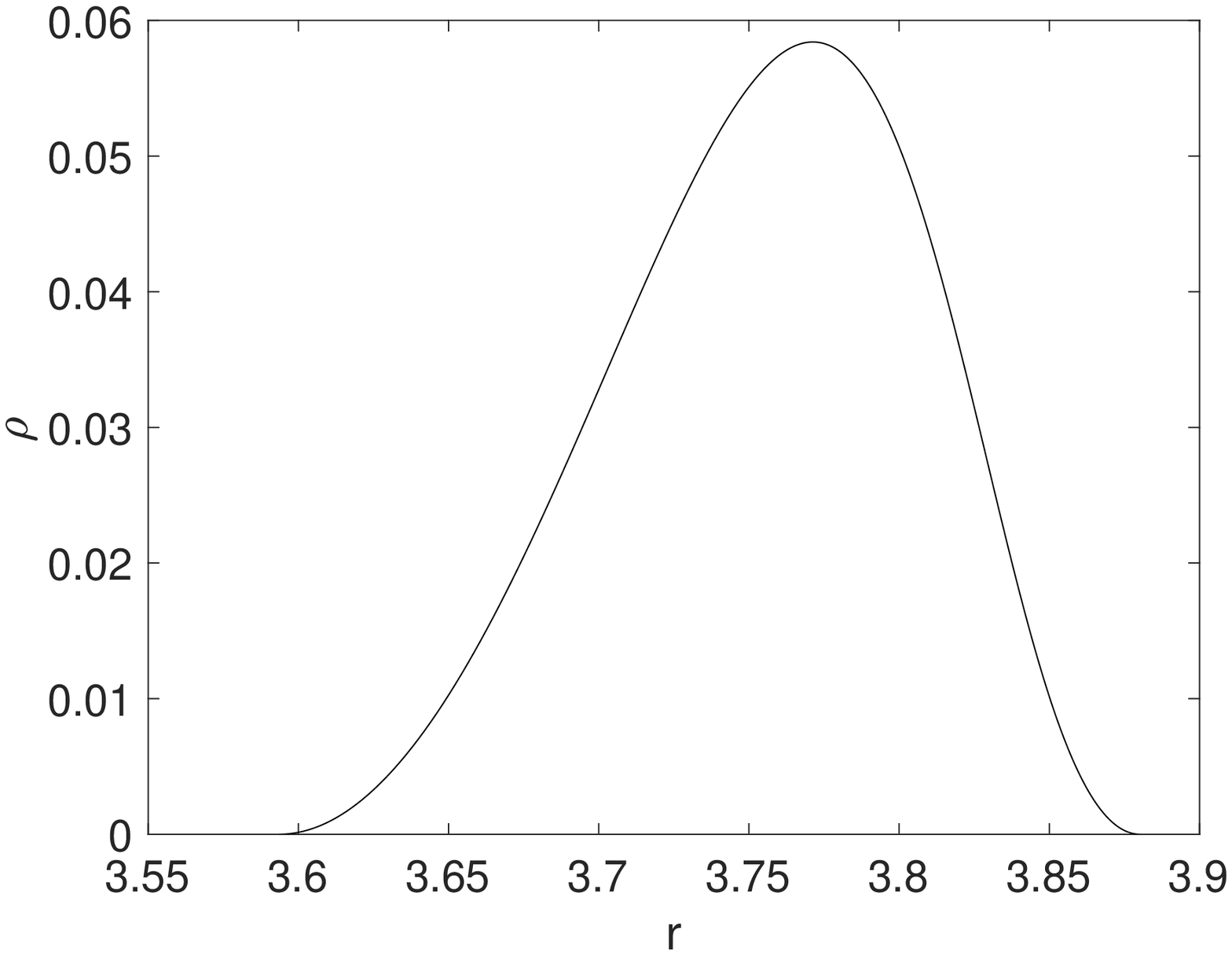}
  \caption{$L_0=15$}
  \label{fig:sfig1ch2}
\end{subfigure}%
\begin{subfigure}{.4\textwidth}
  \centering
  \includegraphics[width=.8\linewidth]{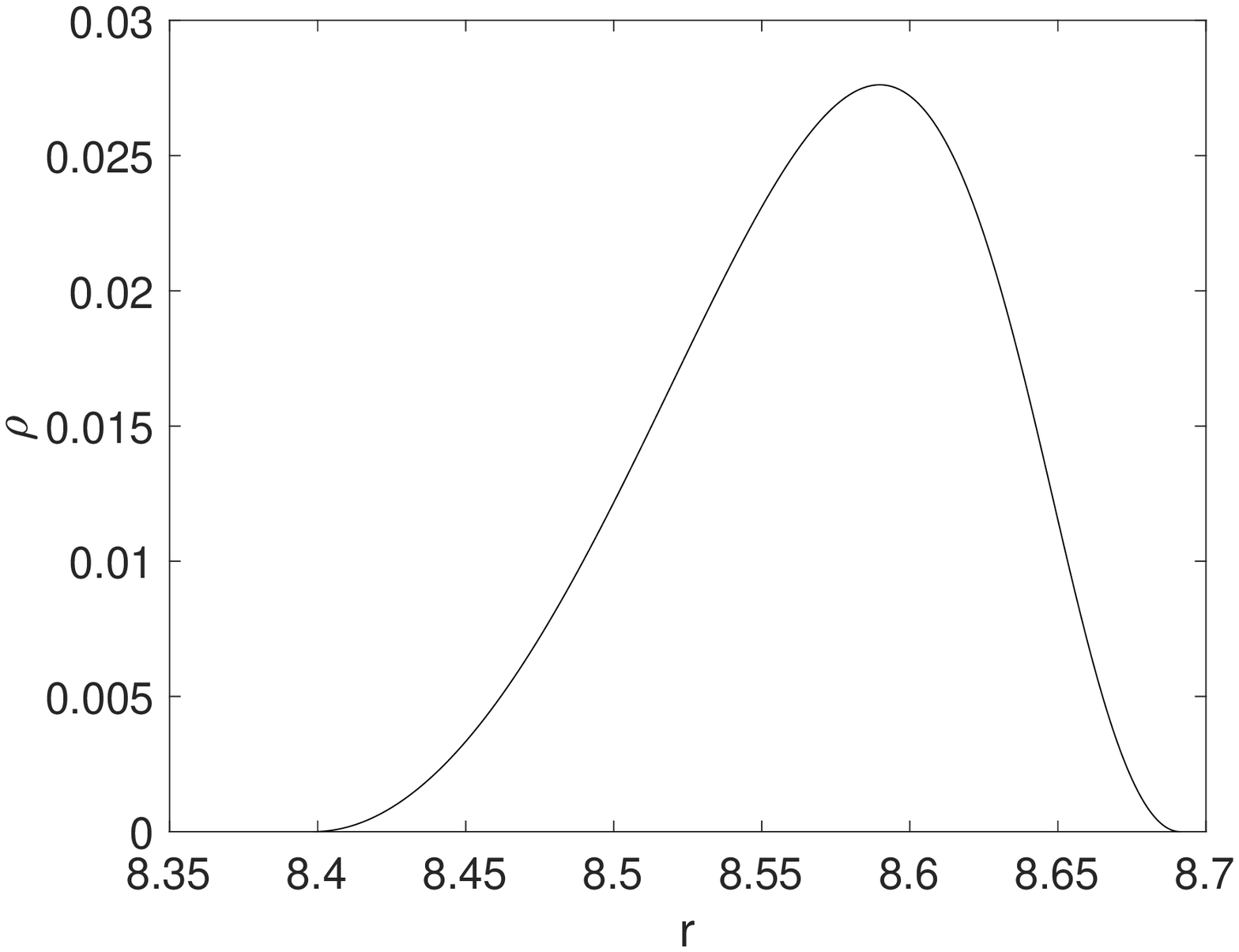}
  \caption{$L_0=75$}
  \label{fig:sfig2ch2}
\end{subfigure}
\begin{subfigure}{.4\textwidth}
  \centering
  \includegraphics[width=.8\linewidth]{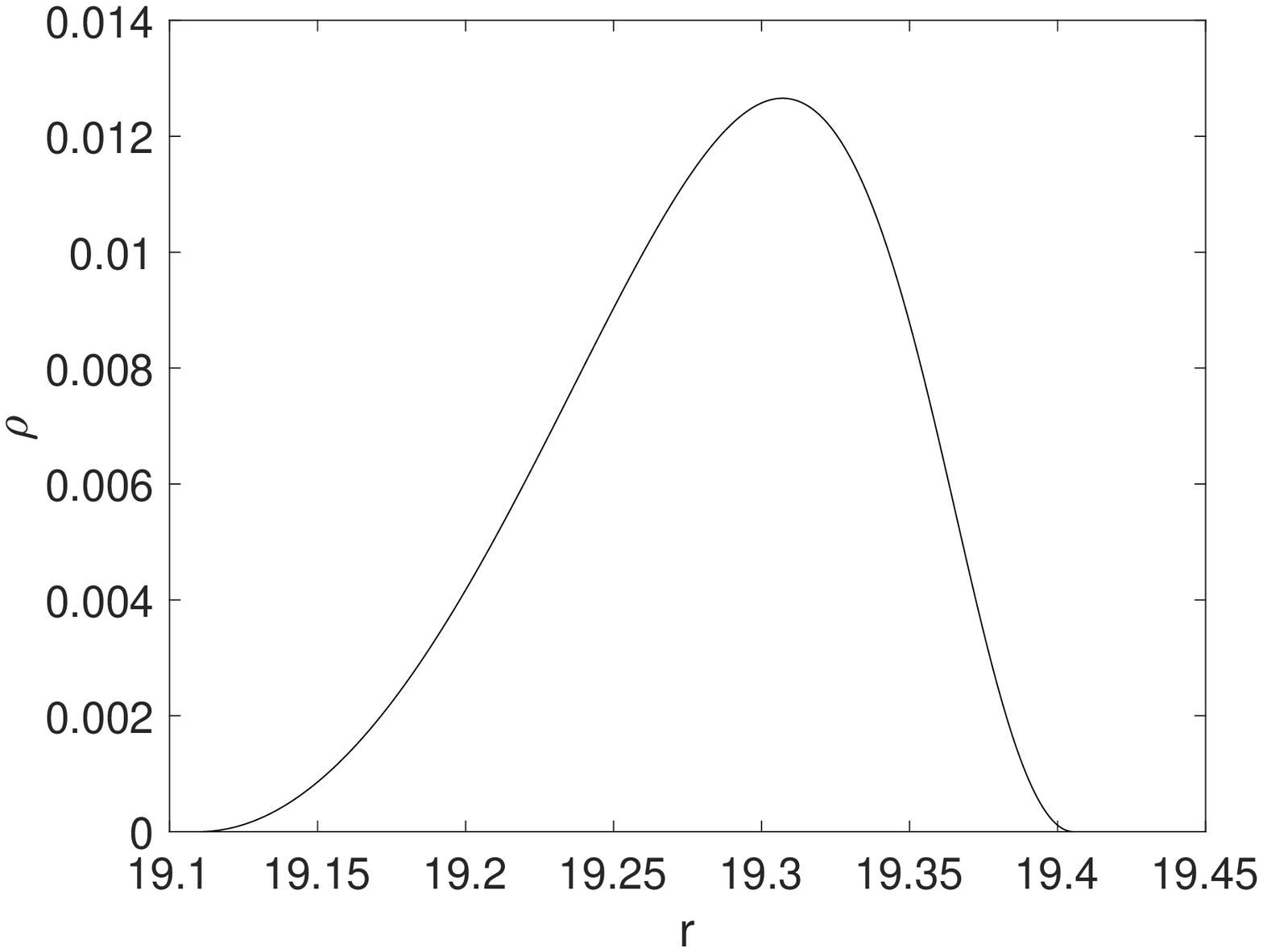}
  \caption{$L_0=375, \Gamma=0.883$}
  \label{fig:sfig3ch2}
\end{subfigure}

\caption{Three shells with constant thickness ($\tau=0$)}
\label{fig:tauzero}
\end{figure}

\begin{figure}
\begin{subfigure}{.4\textwidth}
  \centering
  \includegraphics[width=.8\linewidth]{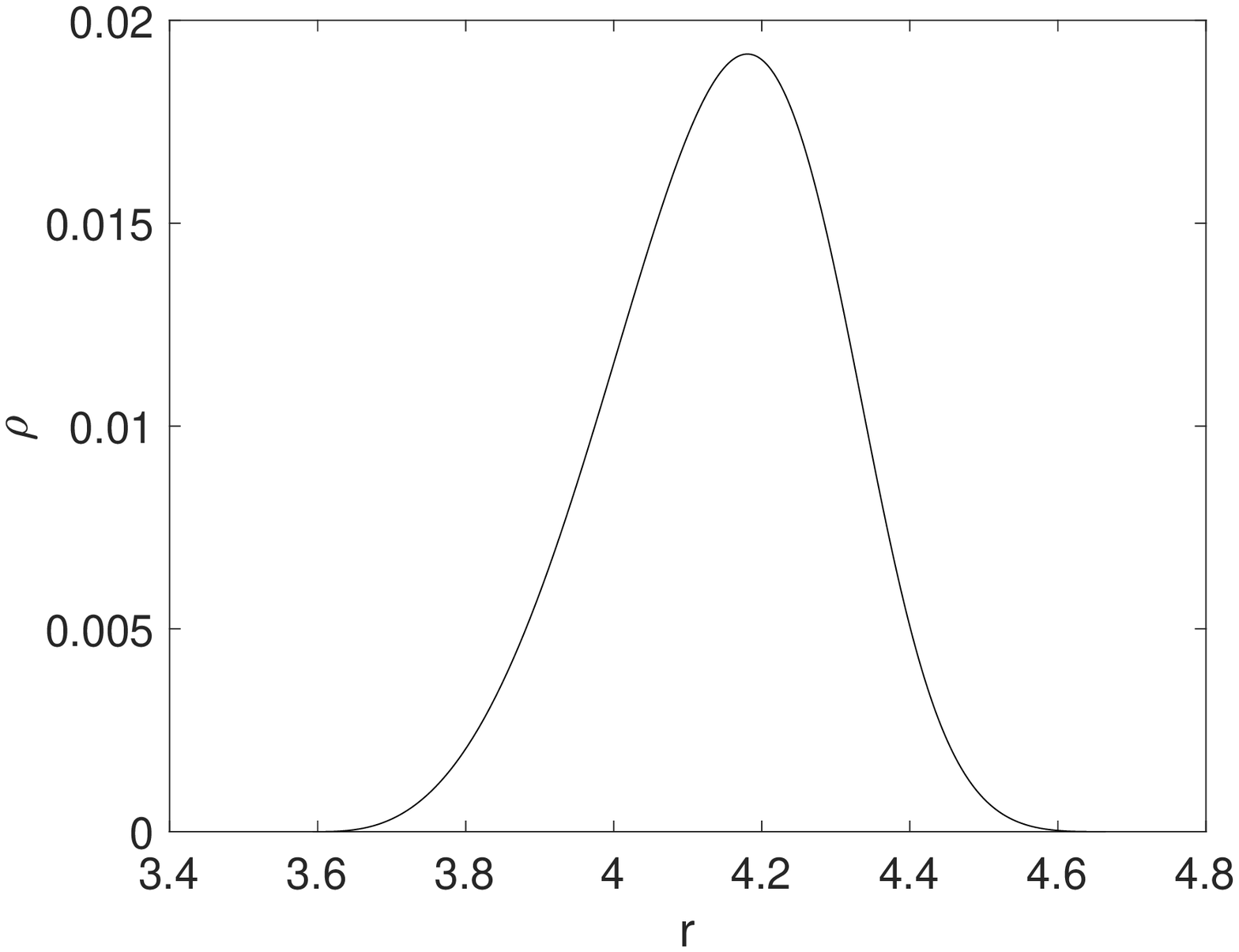}
  \caption{$L_0=15$}
  \label{fig:sfig1ch3}
\end{subfigure}%
\begin{subfigure}{.4\textwidth}
  \centering
  \includegraphics[width=.8\linewidth]{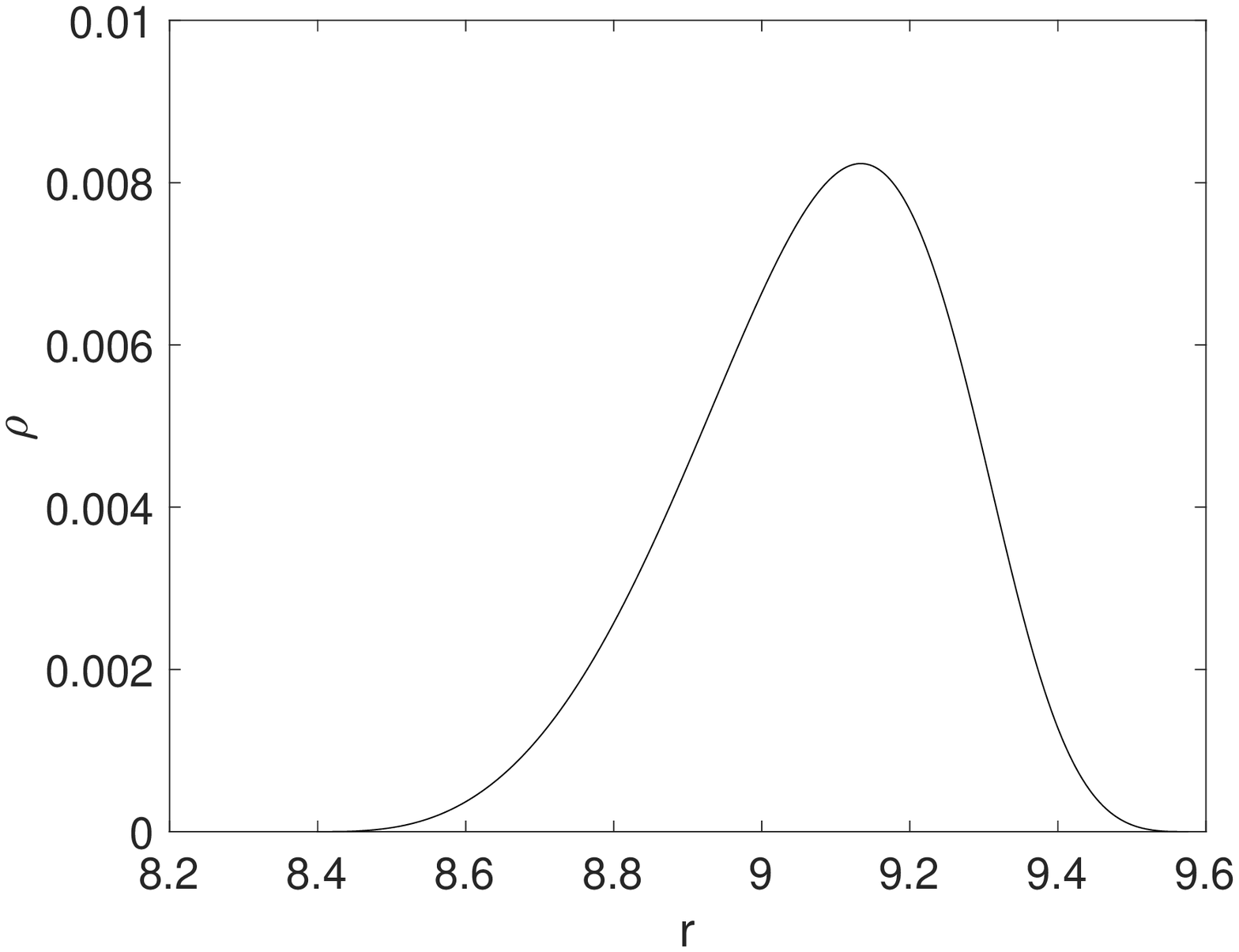}
  \caption{$L_0=75$}
  \label{fig:sfig2ch3}
\end{subfigure}
\begin{subfigure}{.4\textwidth}
  \centering
  \includegraphics[width=.8\linewidth]{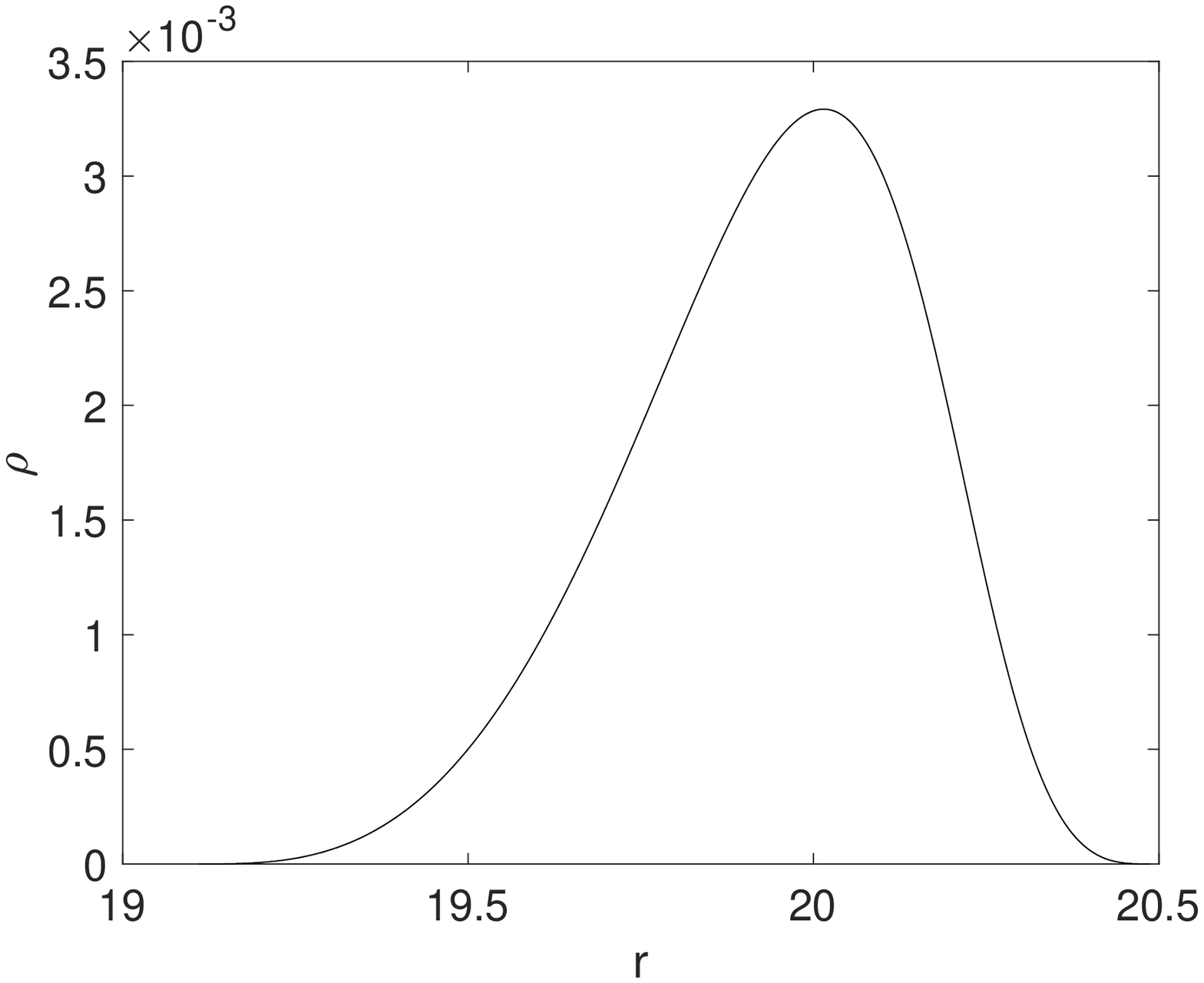}
  \caption{$L_0=375, \Gamma=0.875$}
  \label{fig:sfig3ch3}
\end{subfigure}

\caption{Three shells with growing thickness ($\tau=1/4$)}
\label{fig:tauplus}
\end{figure}

\begin{figure}
\begin{subfigure}{.4\textwidth}
  \centering
  \includegraphics[width=.8\linewidth]{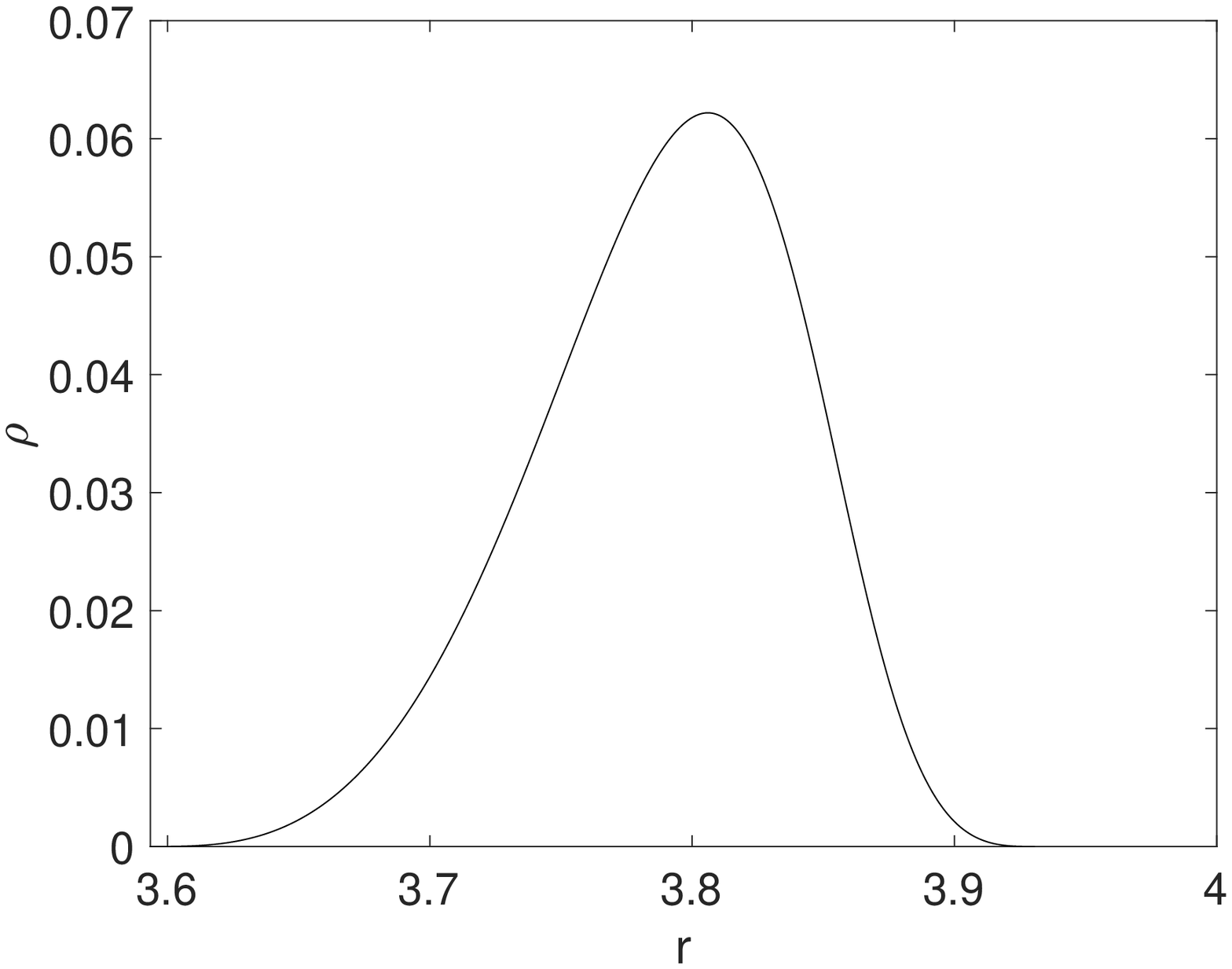}
  \caption{$L_0=15$}
  \label{fig:sfig1ch4}
\end{subfigure}%
\begin{subfigure}{.4\textwidth}
  \centering
  \includegraphics[width=.8\linewidth]{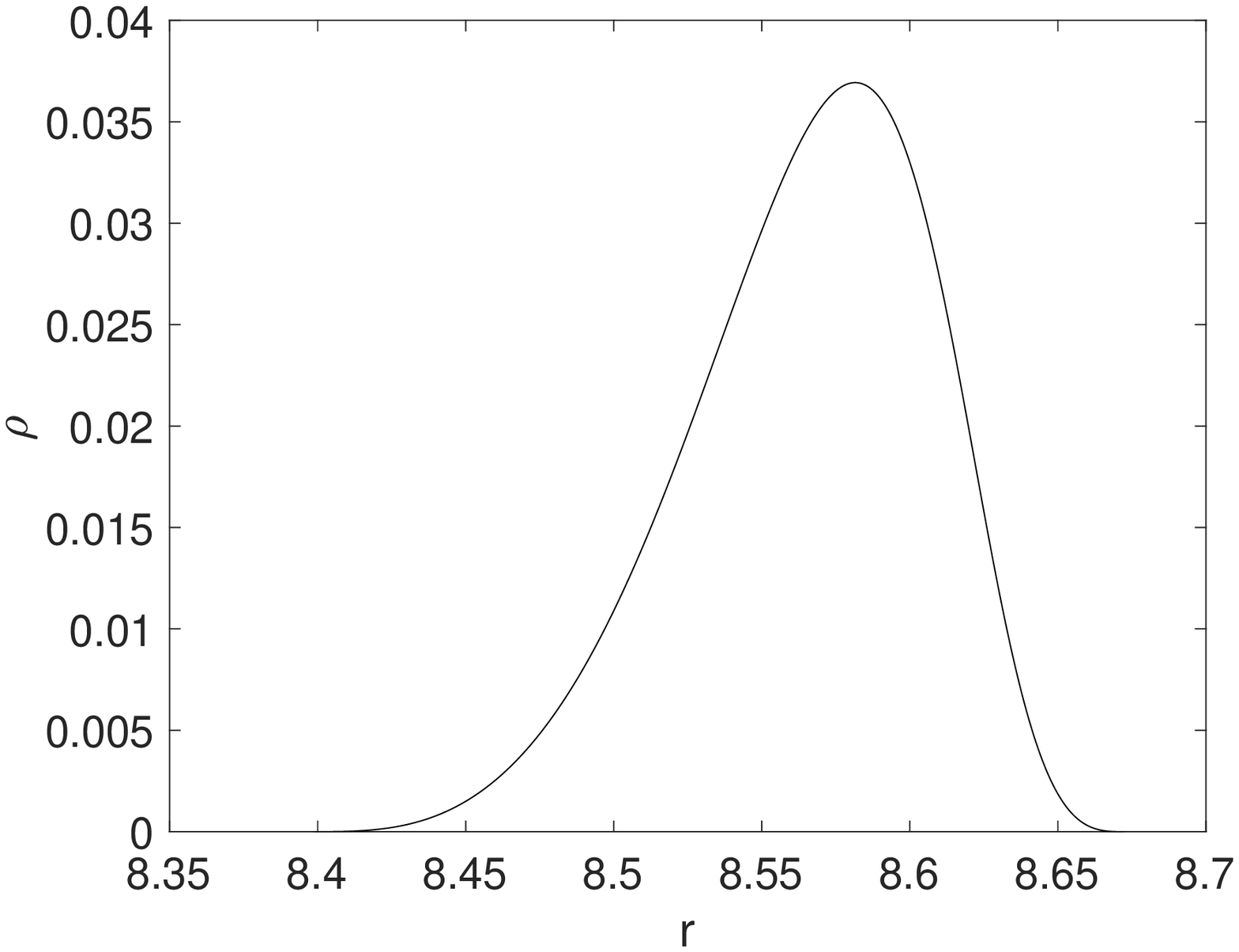}
  \caption{$L_0=75$}
  \label{fig:sfig2ch4}
\end{subfigure}
\begin{subfigure}{.4\textwidth}
  \centering
  \includegraphics[width=.8\linewidth]{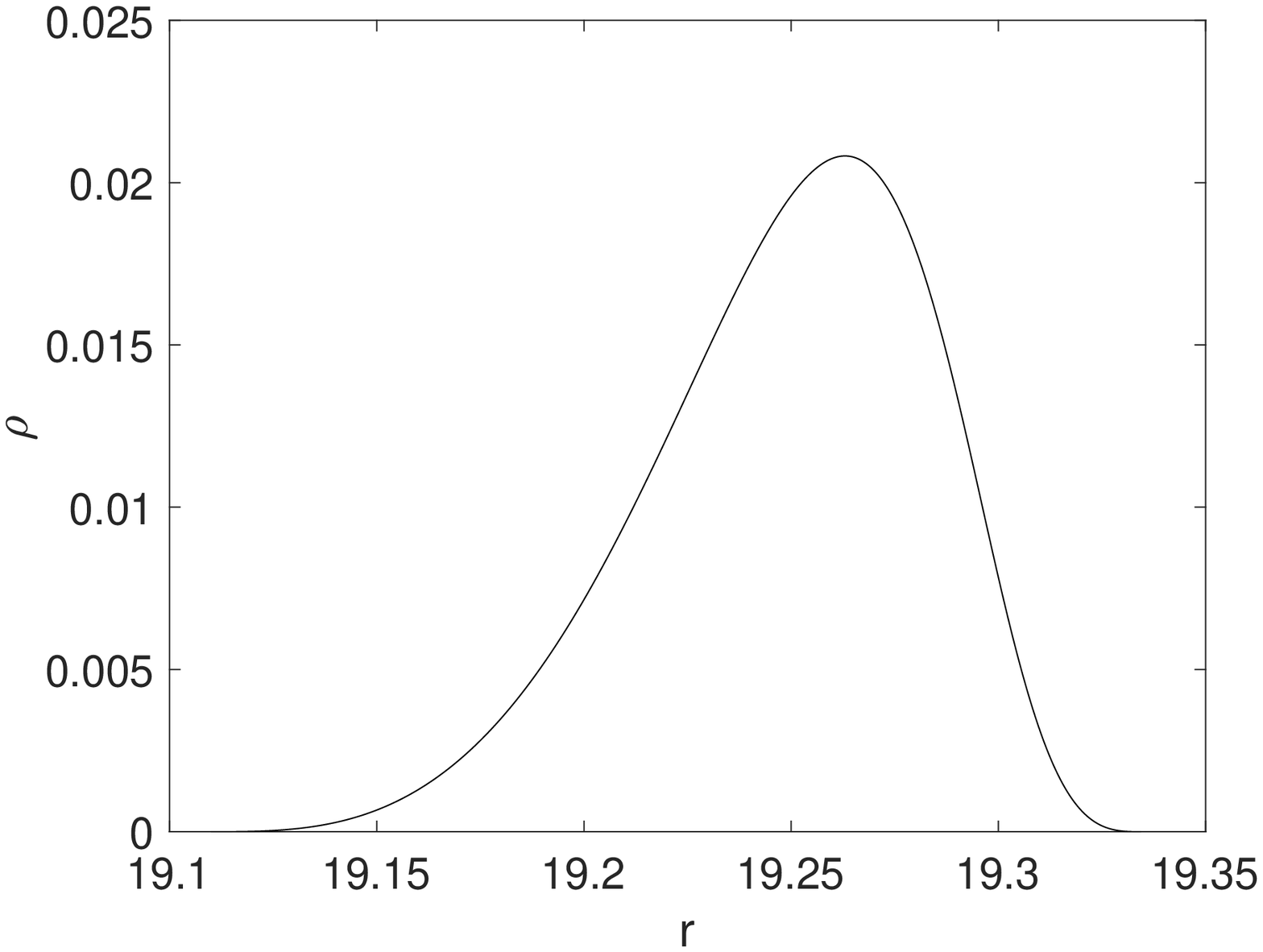}
  \caption{$L_0=375, \Gamma=0.884$}
  \label{fig:sfig3ch4}
\end{subfigure}

\caption{Three shells with decreasing thickness ($\tau=-1/4$)}
\label{fig:tauminus}
\end{figure}

We claim in Remark (\ref{remark-q}) that the thickness of the shells, $R_1-R_0$, depend on the parameters $k$ and $l$ as 
\begin{equation}\label{claimwidth}
R_1-R_0\sim R_0^{\frac{q-2-2l}{q}},
\end{equation}
where $q=k+l+5/2$. In Figure (\ref{fig:tauzero}), (\ref{fig:tauplus}) and (\ref{fig:tauminus}) we have computed shells for $L_0=15$, $L_0=75$ and $L_0=375$ in the three cases $k=0,l=1/2$; $k=1, l=1/2$ and $k=0, l=3/2$. The quantity 
\[
\tau:=\frac{q-2-2l}{q},
\]
takes the following values in these cases, $0, 1/4$ and $-1/4$, respectively. When $\tau=0$, the width of the shell is independent on $R_0$ and this is confirmed in Figure (\ref{fig:tauzero}) where the thickness $R_1-R_0\approx 0.29$ for any of these shells. This is the case we considered in the proof. In  the second case $\tau=1/4$, and the width of the shell thus increases as $R_0$ grows. 
We compute the ratio
\begin{equation}\label{ratiowidth}
\frac{R_1-R_0}{R_0^{\tau}},
\end{equation}
for the three shells in Figure (\ref{fig:tauplus}) and we find that it is approximately constant, roughly $0.67$. In the last case $\tau=-1/4$, and the width decreases as $R_0$ increases. This can be seen in Figure (\ref{fig:tauminus}) where the ratio (\ref{ratiowidth}) is roughly $0.46$ for each shell. Hence, we find numerical support for the claim (\ref{claimwidth}). Moreover, $\Gamma$ is computed in the three cases where $L_0=375$, to confirm that $\Gamma\to\frac89$ as $R_0\to\infty$, cf. Figures \ref{fig:sfig3ch2}, \ref{fig:sfig3ch3} and \ref{fig:sfig3ch4}. 

\subsection{A black hole with several nested shells}
\begin{figure}
\begin{center}
\includegraphics[width=0.4\textwidth]{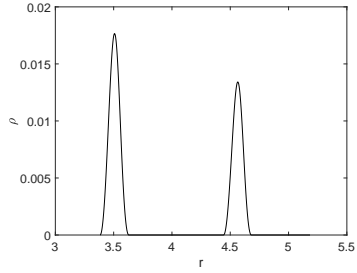}
\end{center}
\caption{$L_0=28.0$ (shell one) and $L_0=48$ (shell two).}\label{rc}
\label{fig2s1}
\end{figure}

Here we construct solutions where the black hole is surrounded by several shells. We choose the black hole mass $M_0=1.0$ so that $L_*=27.0$. We put a shell close to the photon sphere by the choice $L_0=28.0$. We obtain a solution with one shell surrounding the black hole with ADM mass of $1.3$. This implies that $L_*=45.6$ and we choose $L_0=48$ for the second shell. The solution we obtain is depicted in Figure \ref{fig2s1} where we have taken $k=0$ and $l=1/2$. The radius of the photon sphere surrounding the black hole is $r_*=3.0$ and the radius of the photon sphere surrounding the first shell is located at $r_*=4.2$. Similar as above, if we increase the parameter values $k$, then we have to take larger values of $L_0$ to obtain proper shells. Clearly, the procedure can be continued and an arbitrary number of shells can be constructed which surround the black hole. 

\begin{figure}
\begin{center}
\includegraphics[width=0.4\textwidth]{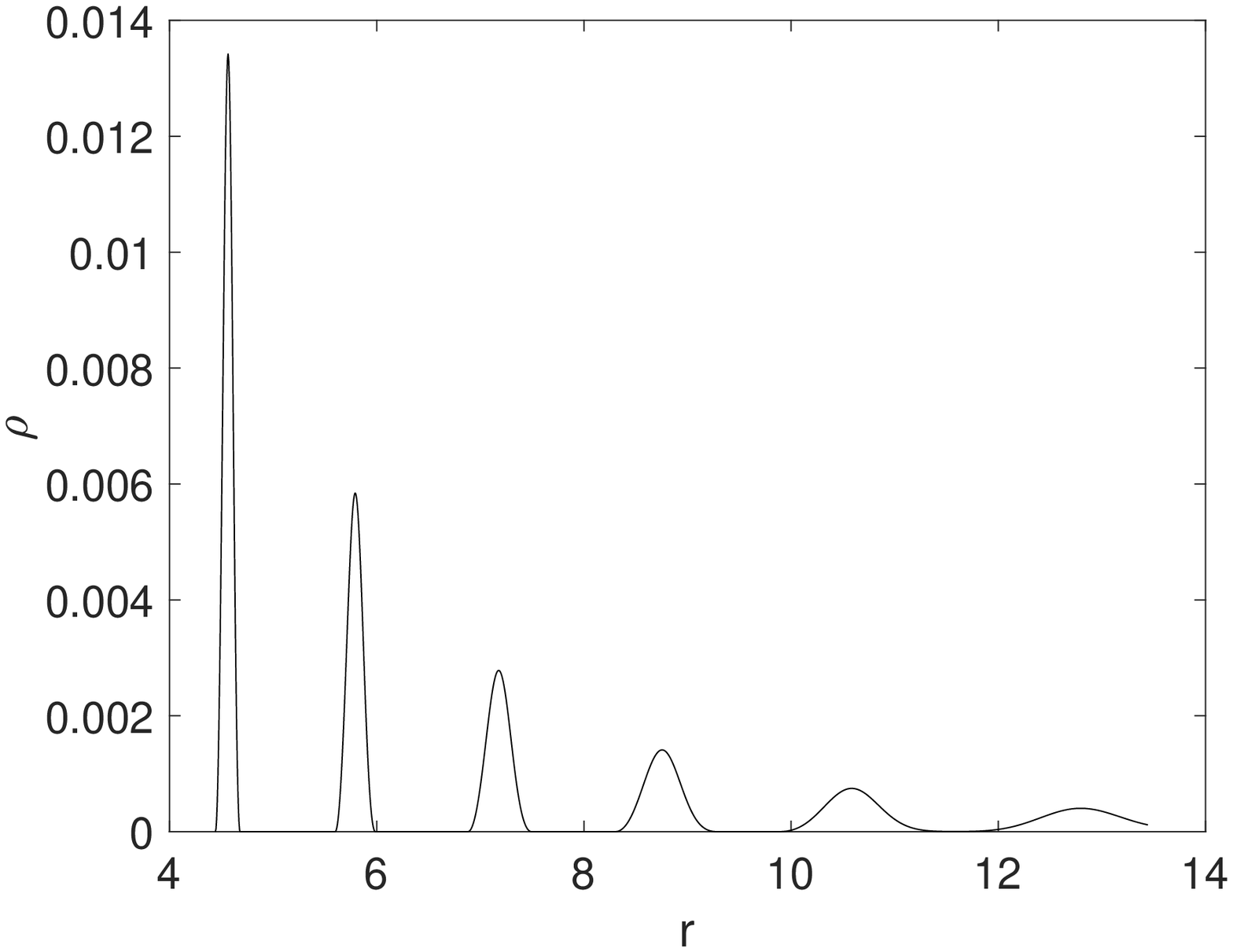}
\end{center}
\caption{Multi-shells with one ansatz. $L_0=48$, $M_0=1.3$, $k=0$ and $l=1/2$.}\label{rc}
\label{figms1}
\end{figure}

\begin{figure}
\begin{center}
\includegraphics[width=0.4\textwidth]{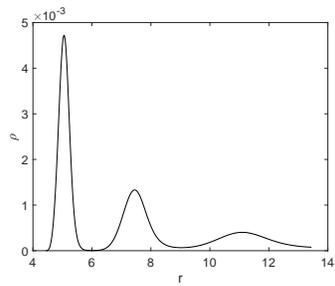}
\end{center}
\caption{No proper shells. $L_0=48$, $M_0=1.3$, $k=0$ and $l=1/2$.}\label{rc}
\label{fignms}
\end{figure}

\begin{figure}
\begin{center}
\includegraphics[width=0.4\textwidth]{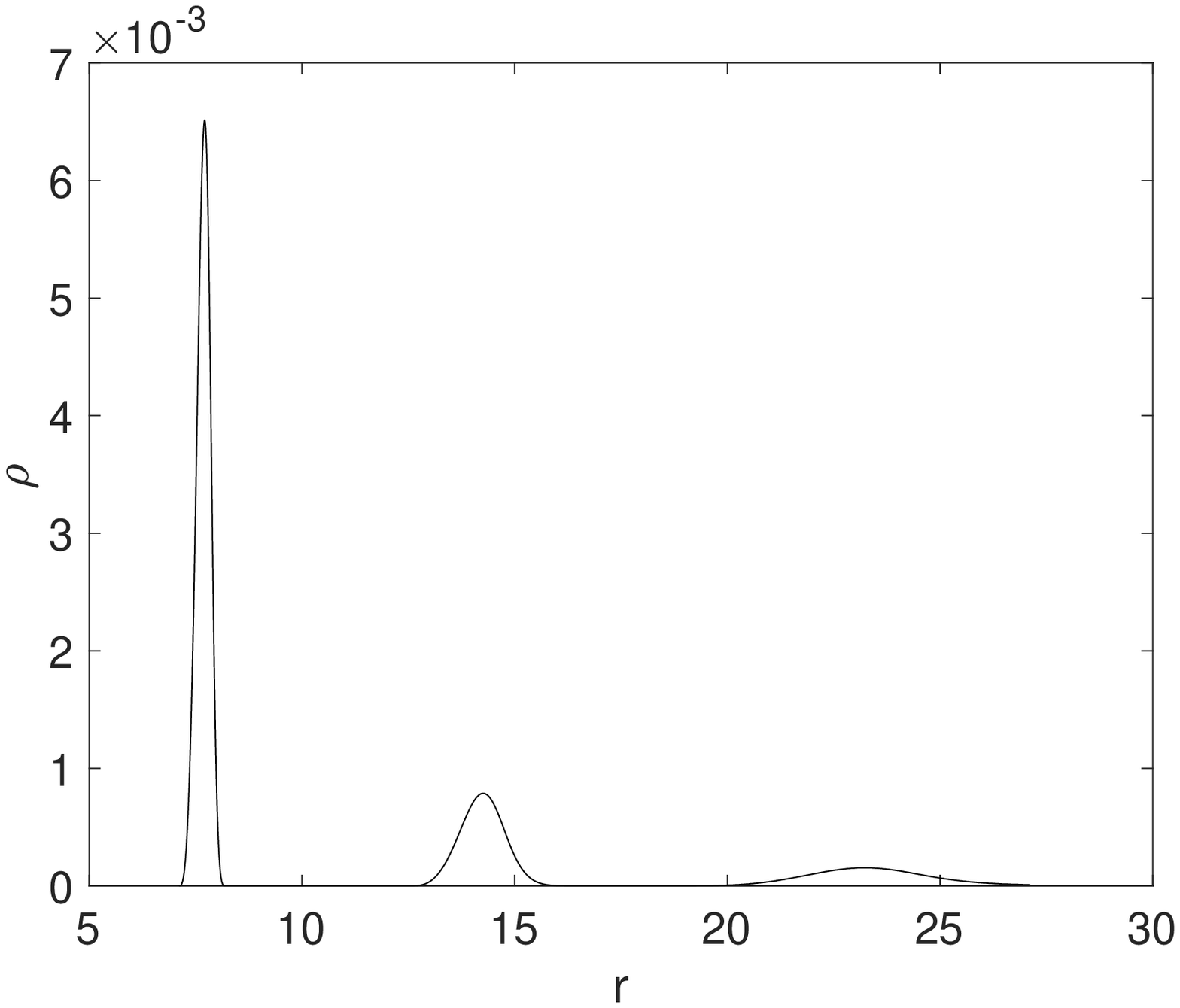}
\end{center}
\caption{Multi-shells with one ansatz. $L_0=80$, $M_0=1.3$, $k=1$ and $l=1/2$.}\label{rc}
\label{figms2}
\end{figure}

\begin{figure}
\begin{center}
\includegraphics[width=0.4\textwidth]{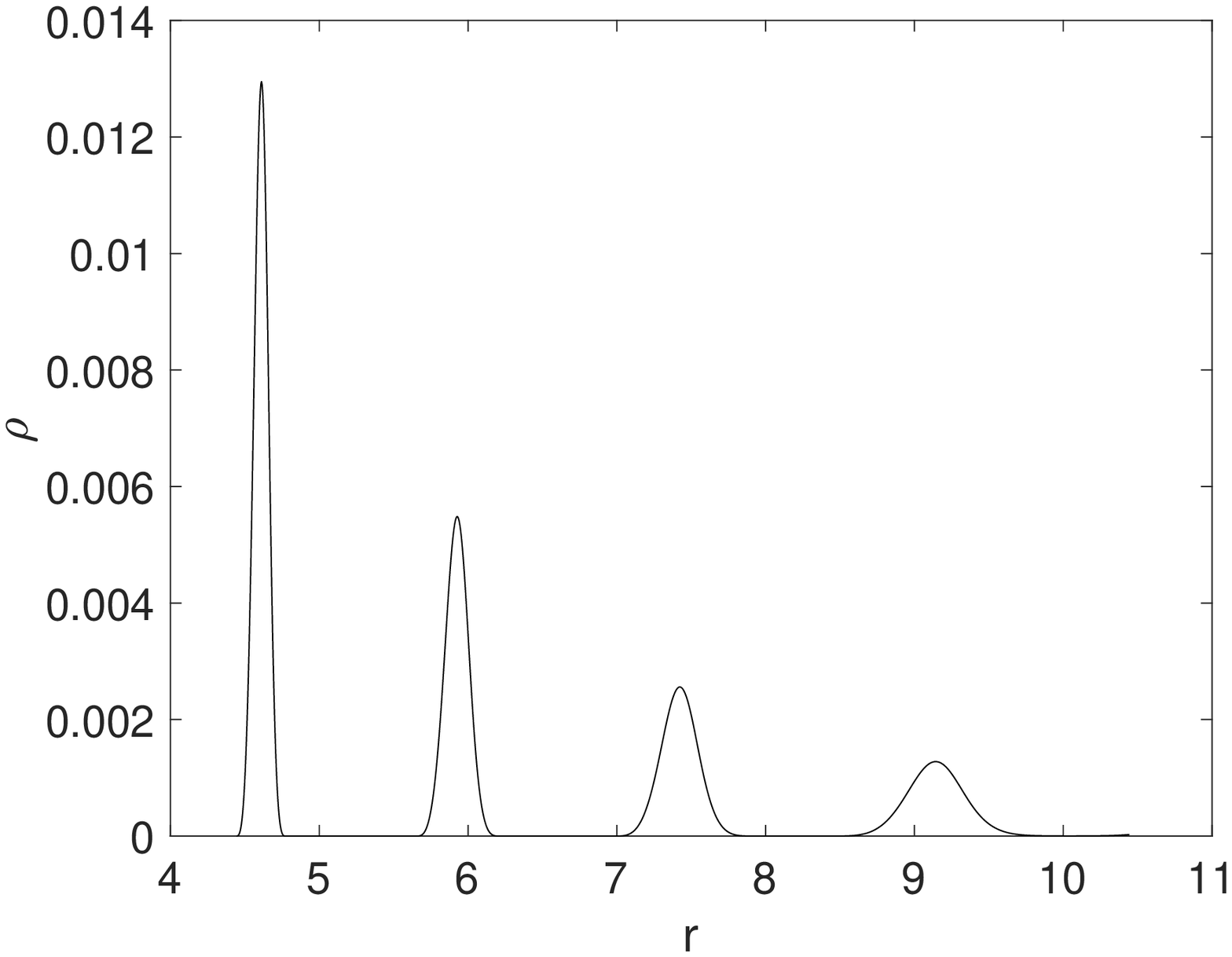}
\end{center}
\caption{Multi-shells with one ansatz. $L_0=48$, $M_0=1.3$, $k=0$ and $l=3/2$.}\label{rc}
\label{figms3}
\end{figure}

Next, we again start with a black hole but we only use \textit{one} ansatz function for the shells, in particular we only solve equation (\ref{groundequation}) once and $L_0$ is fixed. We then obtain a solution consisting of several nested shells, separated by vacuum. 
In Figure \ref{figms1} such a solution is depicted where $M_0=1.3$, $L_0=48$, $k=0$ and $l=1/2$. After the fifth peak there is no longer a vacuum region separating the neighboring shells (which is difficult to see in the picture since $\rho$ is very small) and a Schwarzschild solution has to be glued after the fourth peak in order to obtain an asymptotically flat solution. 
The solution obtained in this way is very similar to the multi-peak solutions obtained in the massive case \cite{AR}. In the latter case there is however no need to glue a Schwarzschild solution since the massive solutions have compact support. We also include a few cases with other choices of the parameters $k$ and $l$. In Figure \ref{fignms} we have used the same parameters as in Figure \ref{figms1} with the only change that $k=1$. In this case there are no proper shells and an asymptotically flat solution is not possible to obtain. By increasing $L_0$ proper shells are obtained as depicted in Figure \ref{figms2}. Finally we consider the case $l=3/2$ but otherwise the same parameters as in Figure \ref{figms1}. In this case three proper shells separated by vacuum are obtained as is depicted in Figure \ref{figms3}. 

\bigskip

\textbf{Acknowledgement: }The author would like to thank Gerhard Rein for useful discussions.

\end{document}